\documentclass[a4paper,fleqn,usenatbib]{mnras}

\usepackage{newtxtext,newtxmath}
\usepackage[T1]{fontenc}
\usepackage{ae,aecompl}

\usepackage{graphicx}	% Including figure files
\usepackage{amsmath}	% Advanced maths commands
\usepackage{csquotes}

\title[LSB galaxy outskirts in HSC]{The nature and origins of the low surface brightness outskirts of massive, central galaxies in Subaru HSC}

\author[Thomas M. Jackson et al.]{Thomas M. Jackson$^{1,2,3}$\thanks{E-mail: tmjackson@tuparev.com},
Anna Pasquali$^{3}$,
Francesco La Barbera$^{4}$,
Surhud More$^{5,6}$,
\newauthor{Eva K. Grebel$^{3}$}
\\
\\
$^{1}$ Tuparev AstroTech, 3 Sofiyiski Geroi Str., Sofia 1612, Bulgaria\\
$^{2}$ Astrosysteme Austria, Galgenau 19, 4212 Neumarkt im M{\"u}hlkreis, Austria\\
$^{3}$ Astronomisches Rechen-Institut, Zentrum f{\"u}r Astronomie der Universit{\"a}t Heidelberg, M{\"o}nchhofstr. 12-14, 69120 Heidelberg, Germany\\
$^{4}$ INAF-Osservatorio Astronomico di Capodimonte, sal. Moiariello 16, Napoli, 80131, Italy\\
$^{5}$ Inter University Centre for Astronomy and Astrophysics, Ganeshkhind, Pune 411007, India\\
$^{6}$ Kavli Institute for the Physics and Mathematics of the Universe (WPI), 5-1-5 Kashiwanoha, Kashiwa 2778583, Japan
}

\date{Accepted XXX. Received YYY; in original form ZZZ}

\pubyear{2022}

\begin{document}
\label{firstpage}
\pagerange{\pageref{firstpage}--\pageref{lastpage}}
\maketitle

\begin{abstract}
We explore the stellar mass density and colour profiles of 118 low redshift, massive, central galaxies, selected to have assembled 90 percent of their stellar mass 6 Gyr ago, finding evidence of the minor merger activity expected to be the driver behind the size growth of quiescent galaxies. We use imaging data in the $g, r, i, z, y$ bands from the Subaru Hyper Suprime-Cam survey and perform SED fitting to construct spatially well-resolved radial profiles in colour and stellar mass surface density. Our visual morphological classification reveals that $\sim 42$ percent of our sample displays tidal features, similar to previous studies, $\sim 43$ percent of the remaining sample display a diffuse stellar halo and only $\sim 14$ percent display no features, down to a limiting $\mu_{r\mathrm{-band}}$ $\sim$ 28 mag arcsec$^{-2}$. We find good agreement between the stacked colour profiles of our sample to those derived from previous studies and an expected smooth, declining stellar mass surface density profile in the central regions (< 3 R$_{\mathrm{e}}$). However, we also see a flattening of the profile ($\Sigma_* \sim 10^{7.5}$ M$_\odot$ kpc$^{-2}$) in the outskirts (up to 10 R$_{\mathrm{e}}$), which is revealed by our method of specifically targeting tidal/accretion features. We find similar levels of tidal features and behaviour in the stellar mass surface density profiles in a younger comparison sample, however a lack of diffuse haloes. We also apply stacking techniques, similar to those in previous studies, finding such procedures wash out tidal features and thereby produces smooth declining profiles. The stellar material in the outskirts contributes on average $\sim 10^{10}$ M$_\odot$ or a few percent of the total stellar mass and has similar colours to SDSS satellites of similar stellar mass.
\end{abstract}


%%%%%%%%%%%%%%%%%%%%%%%%%%%%%%%%%%%%%%%%%%%%%%%%%%

%%%%%%%%%%%%%%%%% BODY OF PAPER %%%%%%%%%%%%%%%%%%

\section{Introduction}
\label{sec:intro}

Early studies of high redshift quiescent galaxies yielded the rather surprising result that massive galaxies at high redshift have significantly smaller effective radii than their low redshift counterparts, despite similar stellar masses \citep{2005ApJ...626..680D}. Numerous follow up studies found similar results \citep{2006MNRAS.373L..36T, 2007MNRAS.382..109T, 2007ApJ...656...66Z, 2008ApJ...677L...5V, 2008A&A...482...21C}. This size difference can amount to a factor of $\sim$ 4 \citep{2007MNRAS.382..109T} and thereby implies stellar density differences of up to 1 or 2 orders of magnitude greater than their low redshift counterparts \citep{2008ApJ...677L...5V}. 

In order to further investigate this size growth, studies utilised samples that span across multiple redshift ranges \citep{2009MNRAS.396.1573F, 2010ApJ...709.1018V}, finding a smooth growth curve across redshift. \citet{2014ApJ...788...28V} built upon these results by extending this study to also include star forming galaxies, finding a similar growth behaviour across redshifts, albeit with a shallower gradient.

Although progenitor bias, which implies that high redshift galaxies may not necessarily be the progenitors of low redshift galaxies \citep{2001ApJ...553...90V}, may be used to argue that these multiple samples are not evolutionarily linked, there are arguments against this. One of these arguments is that there is a lack of compact massive galaxies in the local universe \citep{2009ApJ...692L.118T, 2010ApJ...712..226V, 2010ApJ...720..723T}, which we would expect to observe if all of these objects evolved completely passively and without interactions. A second argument is the old average age of the stellar material in the majority of early type galaxies in the local universe \citep[e.g.][]{2005ApJ...621..673T, 2010MNRAS.404.1775T}, ruling out progenitor bias as the sole driver of these size growth processes. We note, however, that some studies \citep[e.g.][]{2016MNRAS.457.1916D} have shown that some of these high redshift compact objects evolve differently, meaning that the population of present day ellipticals is likely to have multiple evolutionary paths.  

Numerous mechanisms have therefore been suggested to account for this size evolution. Some studies argue that major mergers (stellar mass ratio < 1:4) could increase the size of an object dramatically \citep{2010ApJ...709..218F}. Other studies, however, state that this is likely to increase the stellar mass of an individual object too greatly \citep{2008ApJ...688...48V, 2009ApJ...697.1290B}, especially in the central regions of massive galaxies \citep{2009ApJ...699L.178N}, resulting in much greater numbers of extremely high mass objects than observed in the local universe.

Feedback from Active Galactic Nuclei or AGN is another argument proposed that could cause the size growth of such galaxies \citep[][]{2008ApJ...689L.101F, 2010ApJ...718.1460F}. Numerous counter arguments to this theory, however, have been presented \citep[see][and references therein]{2009ApJ...692L.118T, 2009ApJ...697.1290B}, including that the feedback would need to be extremely fine tuned to reproduce the properties of local, massive ellipticals.

The favoured theory to explain this size growth behaviour is a two phase evolutionary scenario \citep[e.g.][]{2007ApJ...658..710N, 2009ApJ...699L.178N, 2012ApJ...744...63O}. The first phase is a rapid formation of the bulk of the stellar mass. This is followed by the accretion of material from satellites via gas-poor or \enquote{dry} minor mergers (stellar mass ratio $\gtrsim$ 1:4).

Simulations have found a rapid build up of the bulk of the stellar mass via gas accretion, dissipation and rapid, in-situ star formation at high redshifts (from $z = 8$ to $z = 2$). This is followed by an increase in mass and size via minor mergers occurring at lower redshifts, with this secondary phase then connected to the size growth of quiescent galaxies \citep{2007ApJ...658..710N, 2009ApJ...699L.178N}. Similar studies also suggested that massive galaxies are made up of an increasing amount of old ex-situ material towards higher stellar masses and lower redshifts, connecting this ex-situ material to satellite galaxies \citep{2010ApJ...725.2312O, 2012ApJ...744...63O}. Recent simulations have provided more evidence for this scenario, finding that the outskirts of massive galaxies are increasingly dominated by ex-situ material delivered from satellite galaxies \citep{2015MNRAS.449..528H, 2020MNRAS.497...81D}. 

Support for this scenario has also been found in various observational studies. Surface brightness profiles at various redshifts show that the density in the outer regions of galaxies increases the most, likely driven by the accretion of ex-situ material \citep{2009MNRAS.398..898H, 2010ApJ...709.1018V, 2020MNRAS.498.2138B}. Some studies also find a high estimated rate of minor mergers for early type galaxies \citep{2014MNRAS.444..906F}, which is argued to be sufficient to explain the growth of central galaxies below $z \lesssim 1$ \citep{2012ApJ...746..162N}. Arguments such as the expected size growth to mass growth of quiescent galaxies expected by either major merger, AGN feedback and minor merger scenarios have also been invoked in order to favour the minor merger scenario \citep{2009ApJ...697.1290B, 2011MNRAS.415.3903T}, whereby major mergers provide too much mass growth and AGN feedback not enough for the expected size growth. Direct links of accreted stellar material and satellite galaxies in observations are, however, still lacking.

As accreted stellar material from lower mass galaxies is usually extremely diffuse in comparison to the host galaxy, extremely deep imaging is needed in order to detect a signal. Studies ranging from using a number of small aperture telescopes with long exposures \citep[e.g.][]{2010AJ....140..962M, 2017AN....338..503K} to surveys on larger class telescopes \citep[e.g.][]{2012AJ....144..190L, 2013ApJ...765...28A, 2015MNRAS.446..120D, 2018A&A...614A.143M} have been carried out, successfully finding signatures of the accretion of stellar material in a number of different galaxies down to surface brightnesses of $\mu_{g\mathrm{-band}}$ $\sim$ 29 mag arcsec$^{-2}$ in individual galaxies \citep{2020MNRAS.498.2138B, 2020arXiv200713874D}. Simulations also show that as imaging data go deeper, an increasing number of tidal features and distorted stellar material is revealed \citep{2019A&A...632A.122M}. However, a lot of the observational studies considering larger samples, although finding significant signs of tidal material, generally have not attempted to quantify the physical properties such as colour or stellar mass of this material until very recently. This is namely due to a lack of multi-wavelength data that is deep enough. Properties of the material such as colour and stellar mass (as well as the prevalence of this merger activity) can help us understand its origin and thereby possible processes driving the size evolution of galaxies.

In this work we attempt to address this by using deep imaging ($\mu_{r\mathrm{-band}}$ $\sim$ 28 mag arcsec$^2$) from the Hyper Suprime-Cam (HSC) Subaru Strategic Programme (-SSP) of an old, passive galaxy sample in order to search for and quantify signatures of minor merger activity that may be driving the size growth of quiescent galaxies. We firstly morphologically classify our objects in order to quantify the amount of merger activity present and the trends of activity with halo and stellar mass. We then account for PSF effects and implement Voronoi binning \citep{2003MNRAS.342..345C} on the imaging data to attain bins of pixels with high enough signal-to-noise ($S/N$) in order to more accurately constrain fitted Spectral Energy Distributions (hereafter SEDs) using \textsc{cigale} \citep{2009A&A...507.1793N, 2019A&A...622A.103B}. We construct spatially resolved radial profiles in colour from the image photometry and estimates of the stellar mass surface density yielded by the SED fitting in order to investigate and quantify the nature of the tidal features. We repeat these processes on a younger comparison sample. We finally briefly compare our results to other techniques from previous studies on the size growth of massive galaxies.

In Section~\ref{sec:Data} we present the observational data used in this paper. In Section~\ref{sec:Obs_Methods} we outline the methods used to analyse the data. We then present our results in Section~\ref{sec:Results}. In Section~\ref{sec:Discussion} we discuss these results, before summarising our work in Section~\ref{sec:conclusions}.

\section{Hyper Suprime-Cam Data}
\label{sec:Data}

Large scale extragalactic surveys such as the Sloan Digital Sky Survey \citep[SDSS,][]{2000AJ....120.1579Y} have revolutionised our understanding of galaxy evolution by providing a large sample of uniform imaging across a significant area of the sky. The drawback with these samples, however, is that they usually lack the depth of other surveys for more detailed studies such as the Mass Assembly of early-Type GaLAxies with their fine Structures (MATLAS) survey, usually carried out with larger aperture telescopes \citep[][]{2015MNRAS.446..120D, 2020arXiv200713874D}. The disadvantage of such deeper observations however is that the coverage of the sky is usually smaller. The HSC-SSP \citep[][]{2018PASJ...70S...8A} is one survey that is trying to bridge this gap. 

\begin{figure*}
	\centering
	\includegraphics[width=1.6\columnwidth]{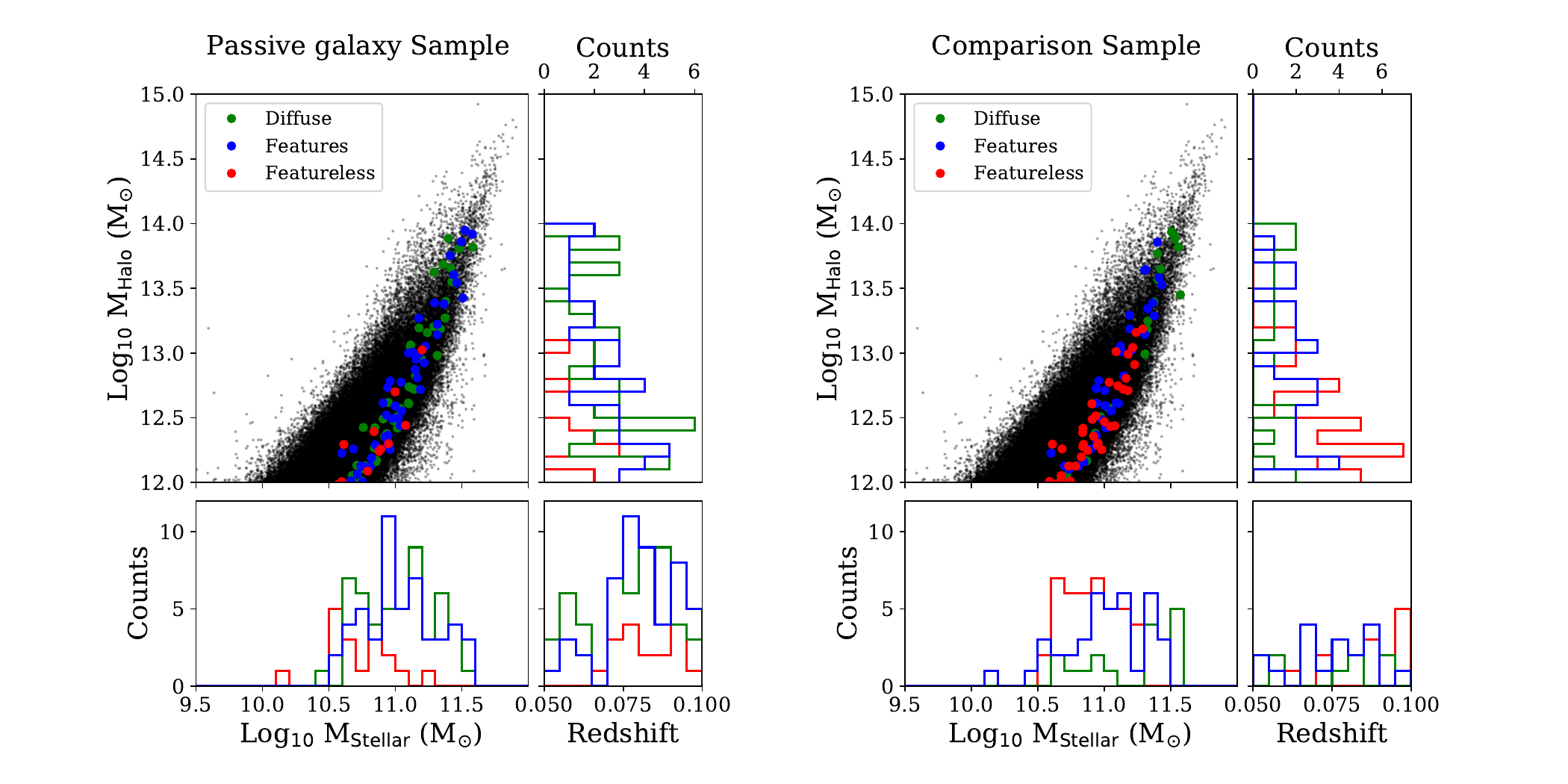}
	\caption{The main panel of each sample shows the distribution of our observational sample in stellar and halo mass phase space. Black points give the parent sample of central galaxies as detailed in \protect\citet{2020MNRAS.497.4262J}. Blue (green, red) points show galaxies with signs of merger activity (a diffuse stellar halo, no activity) as determined by the visual classification described in Section~\ref{OM:classification}. The side panel and bottom panel of each sample show the distributions in stellar and halo mass of each classified sub-sample. The small bottom right panel of each sample shows the distribution of each sub-sample in redshift.}
	\label{fig:sample}
\end{figure*}

The HSC-SSP is an ongoing survey designed to image significant parts of the sky, in 5 different bands ($g, r, i, z, y$). 1400 sq degrees of the sky are being imaged down to 26.1 mag (wide survey), 27 sq degrees down to a magnitude of 27.1 mag (deep survey) and 3.5 sq degrees down to 27.7 mag (ultra-deep survey) in the $r$-band. The camera has a pixel scale of 0.168'' \citep{2018PASJ...70S...1M} and the survey has so far experienced a median seeing in the $i$-band of 0.6'' \citep{2018PASJ...70S...8A}. In order to attain a statistically significant sample we used the wide survey of the incremental data release 2 \citep[300 sq degrees,][]{2019PASJ...71..114A}. The HSC data reduction pipeline \citep{2018PASJ...70S...5B} automatically performs sky subtraction, bias, flat field correction and flux calibration. \citet{2018PASJ...70S...6H} injected synthetic galaxies into HSC images, achieving a 13 and 18 percent precision at a depth of 20 and 25 mag respectively in the {\it i}-band for extended objects when fitting single S{\'e}rsic profiles. The HSC fields have been chosen especially for their low levels of extinction along the line of sight, however we still applied a correction for Galactic foreground reddening to each image according to the law of \citet{1989ApJ...345..245C}, using the extinction maps of \citet{2011ApJ...737..103S}. 

The parent sample is selected from \citet{2020MNRAS.497.4262J}: In this work, we tracked the stellar mass assembly of central galaxies in the SDSS \citep{2000AJ....120.1579Y} as a function of their stellar and halo mass using group catalogues \citep{2017MNRAS.470.2982L}. We used the times at which galaxies assembled 10, 50 and 90 per cent (hereafter $t_{10}$, $t_{50}$ and $t_{90}$), of their stellar mass as determined from the SED fitting techniques of \citet{2012MNRAS.421.2002P, 2016ApJ...824...45P}. These were derived using semi-analytic simulations to generate realistic, non-parametric star formation histories, in conjunction with stellar population models. They were then treated with effects such as dust modelling and nebular emission to generate a library of $\sim$ 1.5 million SEDs to be compared to photometry spanning the UV to NIR. We selected only central galaxies as determined from the friends-of-friends algorithm in the group catalogues of \citet{2017MNRAS.470.2982L}, as they are less affected by environmental processes such as ram pressure stripping compared to satellites. Halo mass cuts of M$_{\mathrm{Halo}} >$ 10$^{12}$ M$_\odot$ were applied in order to select groups and clusters of galaxies.

To obtain the galaxy sample for this study, we applied the following cuts to the parent sample of $\sim$ 90,000 central galaxies from \citet{2020MNRAS.497.4262J}: In order to capture galaxies that have experienced a passive stellar mass evolution over the last few Gyr, and thereby similar systems to those analysed in higher redshift studies, we selected those galaxies contained in the HSC DR2 fields that have a $t_{90}$ value larger than 6 Gyr (reducing the sample to $\sim$ 440 galaxies). We also selected those galaxies with a spectroscopic redshift of 0.05 $< z < $ 0.1 in the group catalogues of \citet{2017MNRAS.470.2982L}, in order to perform the best spatially resolved analysis we could on the individual galaxies themselves, while still retaining a statistically significant sample (reducing the sample to 142 galaxies). We also applied the criterion that each galaxy must have coverage in all 5 bands of HSC ($g, r, i, z, y$, reducing the sample to 134 galaxies) and a minimum of 50 bins at a $S/N > 10$ in its outskirts \citep[beyond the SDSS Petrosian radius of $\sim$ 2 R$_{\mathrm{e}}$ as given in][]{2016ApJ...824...45P} as computed by the Voronoi binning (see Section~\ref{OM:detect_binning}). Selecting galaxies that have a minimum number of the bins in the outskirts may slightly bias our results, whereby galaxies with little activity are likely to have less light/mass excess in their outskirts and are therefore more likely to be discarded, however we stress that only 4 galaxies are discarded from this specific selection process, so we do not expect our results to be significantly impacted. Two galaxies contained in the \citet{2003MNRAS.346.1055K} catalogues defined as AGN or LINER \citep[based on the selection methods of][]{1981PASP...93....5B} were also removed so as to avoid radiation from AGN biasing our results.

In the visual classification we also removed galaxies that had problems in the data with at least 1 band, such as missing data, over-saturation of pixels or imaging artefacts (10 galaxies). The final sample contains 118 centrals from the parent sample of $\sim 90,000$, hereafter referred to as our passive galaxy sample and shown in the left half of Figure~\ref{fig:sample}. The morphological splits are described in Section~\ref{OM:classification} and their relative distributions in halo and stellar mass space are shown in the side and bottom panels of the left half of Figure~\ref{fig:sample}. The bottom right panel of the left half of Figure~\ref{fig:sample} shows the distributions in redshift for later checks in sensitivity biases. 

We also selected a comparison sample of 118 galaxies from the same parent sample of central galaxies using the same selection criteria in order to build a full picture of the evolutionary processes we want to investigate. We selected these galaxies to have as small a combined difference in stellar mass (< 0.1 dex), halo mass (< 0.1 dex) and redshift (< 0.01) phase space to the passive galaxy sample while avoiding selecting the same galaxy twice . This comparison sample was chosen, however, to have a $t_{90} < 4$ Gyr, in order to compare relatively young centrals to our older and passive galaxy sample. This sample is seen in the right half of Figure~\ref{fig:sample}, with the same scheme for the subplots as for the passive galaxy sample in the left half.

We see that each distribution spans a significant range in halo mass ($\sim$ 2 dex) and stellar mass ($\sim$ 1.5 dex). We also notice that those galaxies that display a diffuse stellar halo (see Section~\ref{OM:classification} for more details of the morphological classification) generally occupy the most massive haloes (M$_{\mathrm{Halo}} \gtrsim 10^{13}$ M$_\odot$) and those that display no tidal features (featureless) generally occupy the least massive haloes (M$_{\mathrm{Halo}} \lesssim 10^{13}$ M$_\odot$). The galaxies that display tidal features, indicative of merger activity (features) display a tendency to occupy more massive haloes, however are distributed across the full range of halo masses. These trends are similar in stellar mass, in line with findings from previous studies \citep{2020MNRAS.498.2138B}. \citet{2020MNRAS.498.2138B} find 1.7 times the amount of tidal features above a stellar mass of 10$^{11}$ M$_\odot$ than below, which compares to an increase of 1.2 in our sample (a difference of 1 $\sigma$). This difference may be due to slightly different classification schemes or, more likely, the different data sets, whereby the data of \citet{2020MNRAS.498.2138B} is deeper than that used in this study. They also find an increase in some of their features with increasing environmental density, similar to our results. In the bottom right panel, we see a fairly even distribution of all three sub-samples across redshift. The implications of this behaviour are discussed further in Section~\ref{disc:interpretation}. 

\begin{figure*}
	\centering
	\includegraphics[width=2\columnwidth]{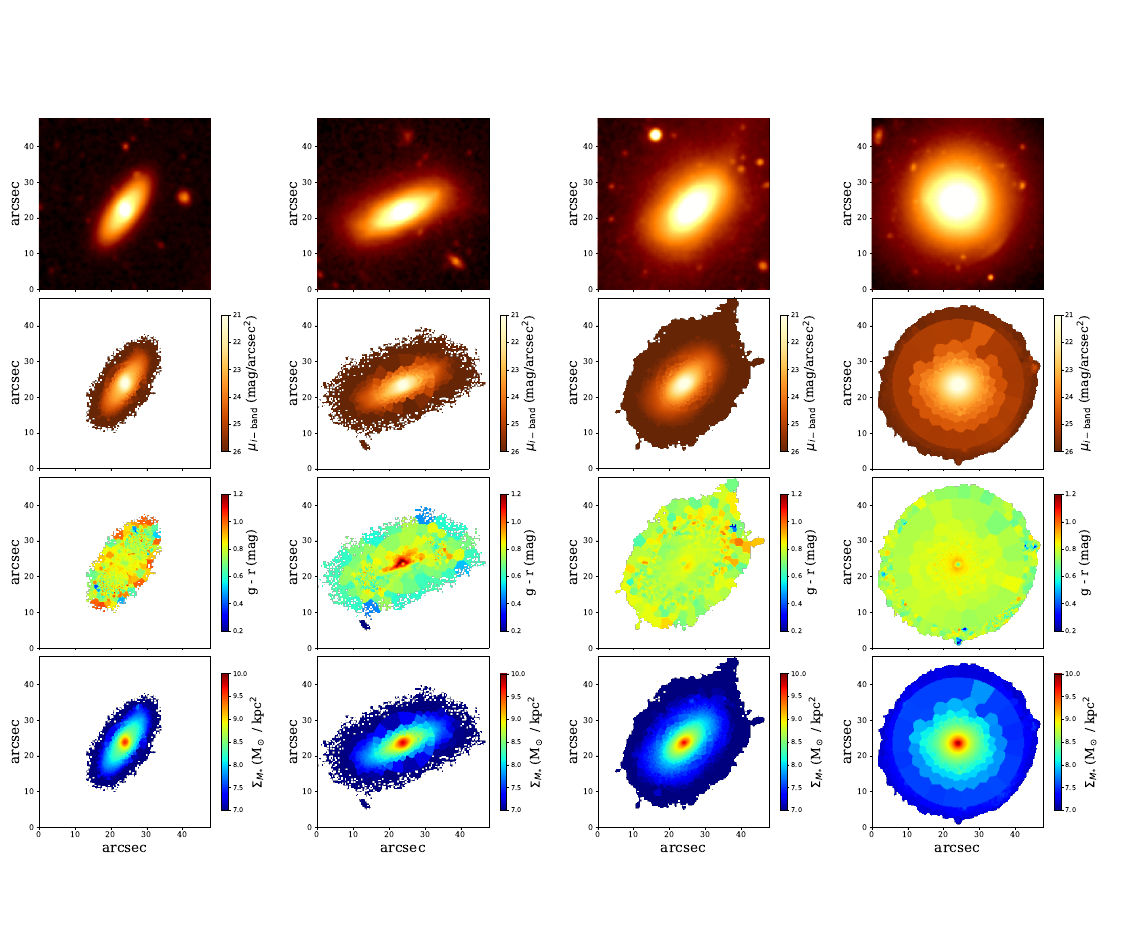}
	\caption{From the top to bottom panels; The {\it i}-band image of a galaxy, the Voronoi binned {\it r}-band surface brightness map, the Voronoi binned {\it g} - {\it r} colour map and the stellar mass density map as yielded by the SED fitting on the Voronoi binned maps. From left to right; an example of a galaxy classified as displaying no merger activity, a galaxy classified as displaying a diffuse stellar halo, a galaxy with a stream of material to the top right and a galaxy that displays a slight shell structure to the bottom right.}
	\label{fig:maps}
\end{figure*}

We also visually classified galaxies into early and late type, finding that $\sim 1$ per cent of our passive galaxy sample is classified as late type with the rest of the sample classified as early type. This statistic rises to $\sim$ 26 per cent for our comparison sample, with $\sim$ 74 per cent of galaxies classified as early type. To reinforce this, we calculated the average S\'{e}rsic index of each sample using the catalogues of \citet{2005AJ....129.2562B} and the bulge to total mass ratios using the catalogues of \citet{2014ApJS..210....3M}. We find that the average S\'{e}rsic index of the passive galaxy sample is 5.00 with an average bulge to total mass ratio of 0.78, compared to an average S\'{e}rsic index of 3.14 for the comparison sample with an average bulge to total mass ratio of 0.60. This shows that our older, passive galaxy sample is dominated by early type morphologies, which have larger S\'{e}rsic indices than the comparison sample.

\section{Observational Methods}
\label{sec:Obs_Methods}

\subsection{Visual classification}
\label{OM:classification}

Visual morphological classification, although with a reasonable degree of subjectivity, has proven to be an extremely useful tool in the study of galaxy evolution, with one of the first and the most used classification systems first developed by \citet{1926ApJ....64..321H}. Morphological classification can be used as an indicator of the current stage of evolution that a galaxy may be at in its lifetime and morphological features or disruptions can help indicate if a galaxy may be undergoing environmental processes such as merging or stripping. Today, researchers can use a number of methods; they can either manually classify their own data set, or if it is too large, public schemes such as Galaxy Zoo \citep{2008MNRAS.389.1179L} have been set up. Machine learning techniques have also been and are being developed for this application \citep[e.g.][]{2010MNRAS.406..342B, 2018MNRAS.473.1108H}. 

As both of our samples are relatively small, each galaxy was independently visually inspected and classified by different co-authors, changing the contrast of each image in the {\it i}-band to search for interaction/accretion features, with a high level of agreement of the classifications between the different co-authors. This technique has already been implemented in previous studies investigating similar types of galaxies, providing us with the advantage of a comparison to previous work \citep{2015MNRAS.446..120D}. We note, however, that there is a degree of subjectivity to this system that needs to be taken into account when drawing conclusions from the results or comparing to previous studies. The {\it i}-band was chosen as it has the best median seeing ($\sim 0.6"$) of all HSC bands. We classified both the passive galaxy sample and the matched comparison sample into three categories. 

In order to quantify the abundance (or lack of) minor merger activity that may be driving the size growth of central galaxies, the first category is those galaxies that display features caused by merger activity. As we are dealing with central galaxies, we expect any features observed to be almost exclusively caused by galaxy interactions with their satellites. This sub-sample was identified in the following ways: those galaxies displaying shells, a ring of material around the main body of the galaxy, streams of material, umbrellas \citep[e.g.][]{2010AJ....140..962M}, those that have clear distortions to the light distribution and those undergoing a minor merger (when a clear and significant secondary peak/core in the light distribution is seen). Some galaxies may display multiple features. This sub-sample is hereafter referred to as "features".

The second category were those galaxies with no clear tidal features but that visually exhibited a diffuse stellar halo (hereafter labelled as diffuse). These galaxies were classified as such as a clear diffuse stellar halo may indicate past merger activity with the accreted material now settled into a state of relative equilibrium. We note that galaxies that have diffuse stellar haloes but also display clear signs of merger activity are classified into the features sub-sample.

The last category is those galaxies that display neither tidal features nor a diffuse stellar halo (hereafter labelled as featureless). Examples of each of the three categories (features, diffuse and featureless) can be seen in Figure~\ref{fig:maps}.

\subsection{Treatment of the imaging data}
\label{OM:detect_binning}

In order to analyse the data while accounting for biases and reducing errors we implemented the following procedures on the imaging data.

As an initial step, we addressed the issue of sky background subtraction: The HSC pipeline performs automatic sky background subtraction, however there is known problem that the pipeline underestimates the sky background in the g- and r-band and overestimates the sky background in the z- and y-band\footnote{\url{https://hsc-release.mtk.nao.ac.jp/doc/index.php/known-problems-2/}}. In order to circumvent this problem, we collected forced photometry (i.e. the coordinates of the apertures used for the photometry are fixed across all bands) on empty parts of the sky from the HSC database performed using a 5 arcsec aperture. We then calculated the median flux per pixel of the sky background of the nearest 100 points to each galaxy in each band. These median flux per pixel values are then subtracted from the imaging data, in order to account for the known systematic problems in the sky background subtraction of the HSC pipeline with the 1 $\sigma$ scatter of the 100 points accounted into the error budget. We note that these median sky background fluxes contribute of the order of between 0.1 and at maximum 5 per cent of the integrated flux of our Voronoi bins, and is smaller than the scatter on the distributions of the integrated fluxes from the Voronoi bins as well as being smaller than the uncertainties of the integrated fluxes in the majority of Voronoi bins.

In order to account for the effects of the Point Spread Function (hereafter PSF) over all bands we measured the Full-Width-Half-Maximum (FWHM) for each galaxy in each band as given by the PSF models constructed by the HSC pipeline \citep[see][for more details]{2018PASJ...70S...5B}. We took the largest FWHM in the entire galaxy sample and convolved each galaxy in each band with the quadrature of the difference between that band and the maximum overall FWHM in order to account for the worst PSF across the entire sample. We note that this technique has been used in the majority of previous studies investigating similar galaxy properties, but that this technique does not take into account the effects of the wings of the PSF. We address this in Section~\ref{res:stacked}. 

We then assigned a threshold $S/N > 3$ per pixel to the imaging data. This signal-to-noise was chosen as lower cuts introduce too much contamination from foreground and background objects, as well as risking sky background fluctuations being introduced into the photometry. On the other hand, higher cuts of e.g. $S/N > 5$ discard many of the tidal features we have visually identified in the data at low surface brightness. This has the effect that the radial profiles we introduce in Section~\ref{res:rad_gradients} do not extend as far and probe the same extent as a $S/N > 3$. We state, however, that taking a higher $S/N$ does not significant change the results of the colour and stellar mass properties of the low surface brightness features in the galacto-centric radii which both $S/N$ cuts cover. We then used the source detection routine from the {\it photoutils} package in {\sc astropy} in order to locate all pixels associated with each galaxy, including the tidal features, at 3 $\sigma$ above the sky background levels (which go down to surface brightness of $\sim$ 28.5 mag arcsec$^{-2}$ per pixel at $S/N$ = 3 per pixel in the g-band). The routine uses an algorithm based on the number of neighbouring pixels also associated with the source to find all pixels associated with a local maximum. Examples of this can be seen in the top two rows of Figure~\ref{fig:maps}, where the top rows show the original {\it i}-band images from HSC and the second rows show the detected galaxies in the {\it r}-band surface brightness maps. 

Some pixels in our sources have low signal-to-noise, however, just reaching the threshold $S/N$ of 3 per pixel. These are usually the low surface brightness features we want to quantify. In order to reduce uncertainties in the estimates yielded by the SED fitting process we used Voronoi binning \citep{2003MNRAS.342..345C} on the images of each galaxy to maximise the signal-to-noise of these features, while still preserving the spatial resolution of higher $S/N$ areas. We took the SDSS Petrosian apertures ($\sim$ 2 R$_{\mathrm{e}}$) for the central areas of our galaxies as this is the aperture used in the study of \citet{2016ApJ...824...45P} from which the parent sample is built in \citet{2020MNRAS.497.4262J} and from within which many of the integrated galaxy properties are estimated such as $t_{90}$. We then Voronoi binned all pixels inside of this aperture to a minimum $S/N$ = 50 per bin, as the signal-to-noise per pixel in these regions is high. For all pixels outside of this aperture, the outskirts of the galaxy, we defined a minimum $S/N$ = 10 per bin, as this is a high enough signal-to-noise to reduce uncertainties in the stellar mass estimates while avoiding the washing out of detail that can occur when Voronoi binning at a signal-to-noise of 50. This results in bins with a $S/N$ >= 10, and with between 1 and $\sim$ 100 pixels, at an average of $\sim$ 4 pixels per Voronoi bin. We also masked bins with contamination from significant foreground and background objects in order to reduce contamination that could bias our results. We note that this makes up only 1.4 percent of all bins, and although increases with radius, due to less bins at higher radii, never contributes above 2.5 percent of all bins, hence we do not expect this to significantly bias our radial profiles. As tidal features are not uniform, we do not attempt to interpolate across neighbouring bins as this method would introduce its own biases to our results.

Total fluxes were then calculated for each Voronoi bin from the imaging data in the following manner: The median flux of each Voronoi bin was calculated and multiplied by the amount of pixels in the Voronoi bin, so as to avoid fluctuations due to contaminating objects that can bias the mean, especially at low surface brightness. Associated errors were calculated using the variance maps added in quadrature with the flux calibration error. The fluxes and associated errors were then converted into magnitudes for colour profiles, with an median error of $\sim$ 0.2 mag, smaller than the scatter in the stacked radial colour profiles of both samples of galaxies. We also calculate the physical area of each Voronoi bin, which ranges between 0.037 and $\sim$ 50 kpc$^{2}$, allowing us to calculate the stellar mass surface density after the SED fitting process outlined in Section~\ref{OM:CIGALE} and the surface brightness per bin, finding a minimum of $\sim$ 28 mag arcsec$^{-2}$ per bin at a $S/N$ of 10 per bin. We also ran Voronoi binning at constant signal-to-noise ratios of 20 and 30 per bin, to verify if our adaptive Voronoi binning process introduces biases to our results, finding no significant differences in the profiles, however the loss of spatial resolution means fewer Voronoi bins, and hence more stochastic fluctuations and profiles which are not as extended.

The results of this process can be seen in Figure~\ref{fig:maps}, where the third panel shows the corresponding {\it g} - {\it r} colour map and the bottom panel shows the corresponding stellar mass surface density map as recovered by the SED fitting process described in Section~\ref{OM:CIGALE}. From left to right we see an example of a featureless galaxy, a diffuse galaxy, a galaxy that exhibits a stream and a galaxy that displays a shell (both classified in the features sub-sample).

\subsection{SED fitting with \textsc{cigale}}
\label{OM:CIGALE}

In order to estimate the stellar mass corresponding to each Voronoi bin in our data, we used the SED fitting software \textsc{cigale} \citep{2005MNRAS.360.1413B, 2009A&A...507.1793N, 2019A&A...622A.103B}. We chose \textsc{cigale} due to its speed, since due to the resolution and depth of our imaging data, the Voronoi binning process yielded over 500,000 SEDs that needed fitting for each sample. \textsc{cigale} can fit SEDs spanning from the Ultraviolet (UV) to the Infrared (IR) range of the electromagnetic spectrum and is based on an energy balance principle, whereby the attenuation due to dust of the UV radiation produced by massive young stars is expected to be re-radiated in the IR part of the spectrum. It also has models to account for AGN emission, dust attenuation, multiple star formation histories and stellar population models. 

As we expected these systems to be dominated by old stellar material, however with the possibility of some recent star formation triggered by interactions, we generated models in \textsc{cigale} based on double-exponential star formation histories spanning ranges between 2 and 12 Gyr for the main burst. As our passive galaxies are selected to have formed 90 percent of their stellar mass over 6 Gyr ago (this is between 3 and 4 Gyr for our comparison sample in almost all cases), we limit the additional burst to a contribution between 0.1 and 5 per cent of the mass with ages ranging between 100 Myr and 2 Gyr. The best fit SFH contains a 0.1 percent secondary burst in the overwhelming majority of cases. We then used these star formation histories with the synthetic stellar libraries of \citet{2003MNRAS.344.1000B}, assuming a \citet{2003PASP..115..763C} initial mass function and metallicites spanning the full range from 1/30$^{\mathrm{th}}$ solar to 4 times solar in order to generate the model SEDs to fit to the HSC photometry. We neglected AGN contributions to the modelling as we removed AGN from the sample as described in Section~\ref{sec:Data}.

We also inputted the {\it g}, {\it r}, {\it i}, {\it z} and {\it y} HSC filter functions into \textsc{cigale} and fit the SEDs. The average uncertainties yielded by the SED fitting process for the estimations of stellar masses are of the order of 0.3 dex, which when accounted for in the stellar mass surface densities is, on average, smaller than the scatter in the corresponding radial profiles of our sample of galaxies, meaning the errors do not significantly bias our results.

\section{Results}
\label{sec:Results}

\subsection{Levels of merger activity}
\label{res:activity}

\begin{figure*}
	\centering
	\includegraphics[width=2\columnwidth]{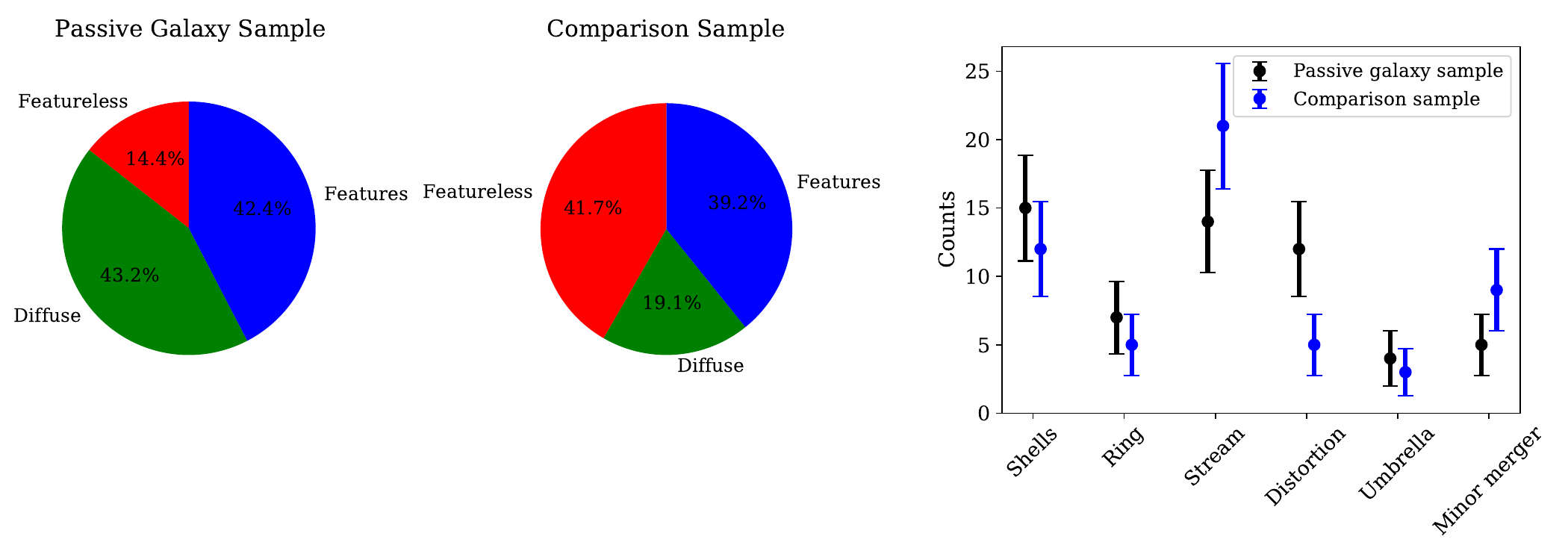}
	\caption{The left hand panel shows a pie chart of the percentage of the passive galaxy sample that was classified as either displaying visual signs of merger activity (features), no activity but a diffuse stellar halo (diffuse), or no activity at all (featureless). The central panel shows the same as the left hand panel but for our younger comparison sample. The right hand panel shows the sub-samples of galaxies that were classified as displaying visual signs of merger activity, split dependent on the type of merger activity.}
	\label{fig:split_stats}
\end{figure*}

Figure~\ref{fig:split_stats} shows the statistics yielded by our morphological classification. The left hand panel shows the split of our passive galaxy sample into those centrals that display tidal features (features), which make up 42.4 per cent of the sample. We note that this percentage is very similar to previous studies such as \citet{2015MNRAS.446..120D}. Those that show a diffuse stellar halo (diffuse) make up 43.2 per cent of our sample and those that are featureless contribute 14.4 per cent of our passive galaxy sample. 

When we compare these statistics to the comparison sample of younger central galaxies in the central panel we see that the percentage of centrals classified with features remains similar at 39.2 per cent. The percentage of galaxies classified as diffuse, however, drops to 19.1 per cent and the percentage of galaxies classified as featureless increases to 41.7 per cent. Assuming Poissonian errors on each sub-sample, we can calculate that the differences in the diffuse and featureless sub-samples seen between the passive and comparison samples are significant (> 5 $\sigma$), however are not significant (< 1 $\sigma$) in those galaxies classified as exhibiting features. We discuss these differences and possible scenarios for this behaviour in Section~\ref{disc:interpretation}.

The right hand panel shows the further morphological classification we used on the features sub-sample of galaxies. We note that some galaxies may display different kinds of features simultaneously and therefore may count many times in this specific sub-plot. We see a range of different features, from shells and significant distortions in the light distribution to minor mergers and rings. In our passive galaxy sample we see that distortions and shells are the most prevalent signs of merger activity with respect to the comparison sample. The comparison sample, on the other hand, shows an increased amount of streams compared to our passive galaxy sample, with all three signs of merger activity the most prevalent in both galaxy samples. When we calculate the Poissonian errors, we find no significant differences (< 1 $\sigma$) in the number of shells, rings and umbrellas, differences above 1 $\sigma$  n the number of streams and minor mergers and differences above 2 $\sigma$ in the numbers of distortions. We caution, however, against over interpretation due to relatively small number statistics.

\subsection{Radial Profiles}
\label{res:rad_gradients}

One of the most effective ways to investigate the star formation and assembly histories of galaxies is to plot radial profiles of various properties. This can help reveal possible evolutionary scenarios which a galaxy has undergone. For example metallicity and age profiles in the central regions may reveal inside out star formation \citep[e.g.][]{2011ApJ...740L..41L}. The outskirts, however are dominated by ex-situ stars according to simulations \citep{2020MNRAS.497...81D}, hence profiles out to large R$_\mathrm{e}$ reveal the nature of accreted material \citep{2012MNRAS.426.2300L, 2015MNRAS.449..528H, 2020MNRAS.497...81D}, which is the aim of this study. We therefore utilise this technique to explore the nature of the colour and stellar mass surface density of the accreted material in both galaxy samples. 

From the Voronoi binning process we calculated an average distance for each bin to the centre of the galaxy. Using this we then stacked every Voronoi bin from every galaxy in our passive galaxy sample in bins of 0.1 R/R$_\mathrm{e}$ and constructed distributions of {\it g} - {\it r} colour and stellar mass surface density. We use the effective radius as measured by SDSS (hereafter R$_\mathrm{e}$), in order to be able to better compare these results to previous studies. We note that when we measure the light profiles from our HSC data, we find the majority of our measured effective radii and those measured from SDSS have a difference of less than 10\%, which is in line with previous studies \citep{2015MNRAS.446..120D} and which we do not expect to significantly bias the results presented here. We also add the constraint that each bin of 0.1 R/R$_\mathrm{e}$ requires 100 Voronoi bins, to reduce statistical biases due to stochastic fluctuations. We note that very few Voronoi bins (< 0.1 percent) contains pixels in multiple radii. From the distribution in each bin we calculate the median of each bin, and the 5$^{\mathrm{th}}$, 16$^{\mathrm{th}}$, 25$^{\mathrm{th}}$, 75$^{\mathrm{th}}$, 84$^{\mathrm{th}}$ and 95$^{\mathrm{th}}$ percentiles to construct our stacked radial profiles. We also normalise each bin of stellar mass density by the total stellar mass within R$_\mathrm{e}$ of each galaxy following previous studies (we note that normalising by 2 or 3 R$_\mathrm{e}$ has a negligible difference on the results).  

\begin{figure*}
	\centering
	\includegraphics[width=1.9\columnwidth]{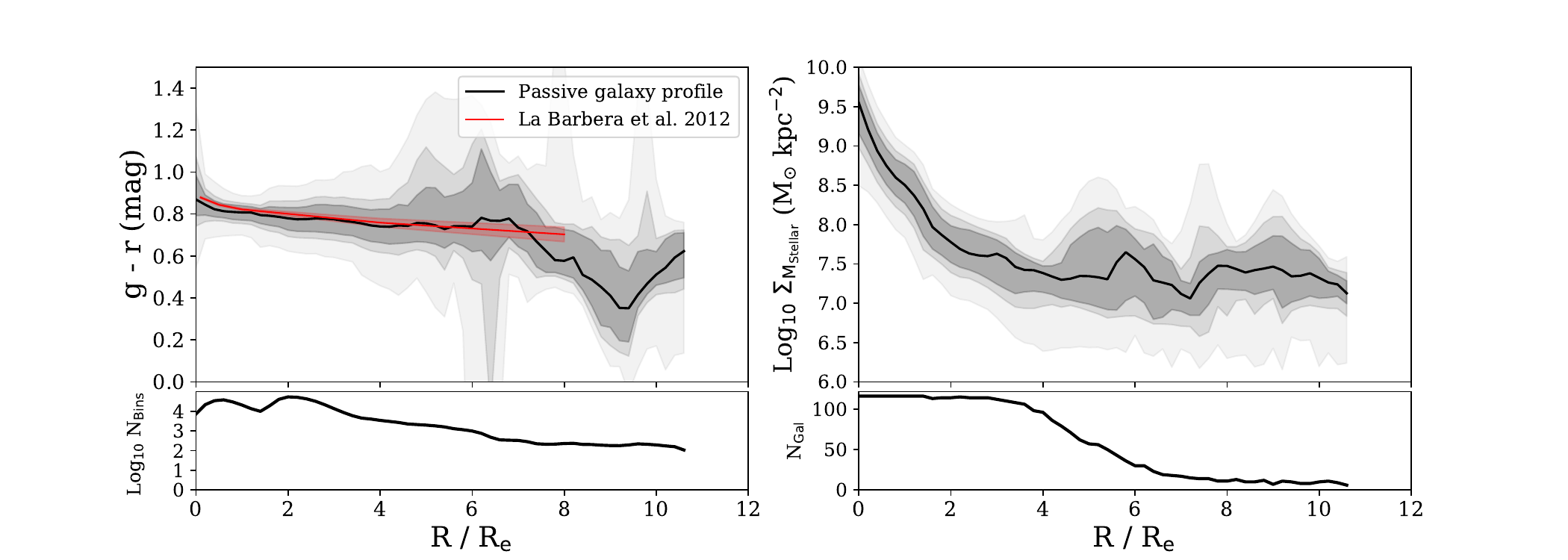}
	\caption{The top left hand panel displays the stacked radial profile (in bins of 0.1 R/R$_\mathrm{e}$) of {\it g} - {\it r} colour for the 118 galaxies in our older ($t_{90} > 6$ Gyr), passive galaxy sample. The solid black line gives the median of the distribution of each bin, while the shaded areas represent the 5$^{\mathrm{th}}$, 16$^{\mathrm{th}}$, 25$^{\mathrm{th}}$, 75$^{\mathrm{th}}$, 84$^{\mathrm{th}}$ and 95$^{\mathrm{th}}$ percentiles of each stacked bin. For comparison, the results from the study of \protect{\citet{2012MNRAS.426.2300L}} are given by the solid red line and shaded areas. The top right hand panel shows the stacked radial profile (also in bins of 0.1 R/R$_\mathrm{e}$) of the stellar mass surface density with corresponding percentiles, normalised by the integrated stellar mass within 1 R$_\mathrm{e}$. The bottom left and bottom right panels show the log of the number of Voronoi bins per bin in radius and the number of galaxies contributing to the profile within any one bin in radius respectively.}
	\label{fig:gr_radial}
\end{figure*}

\begin{figure*}
	\centering
	\includegraphics[width=1.9\columnwidth]{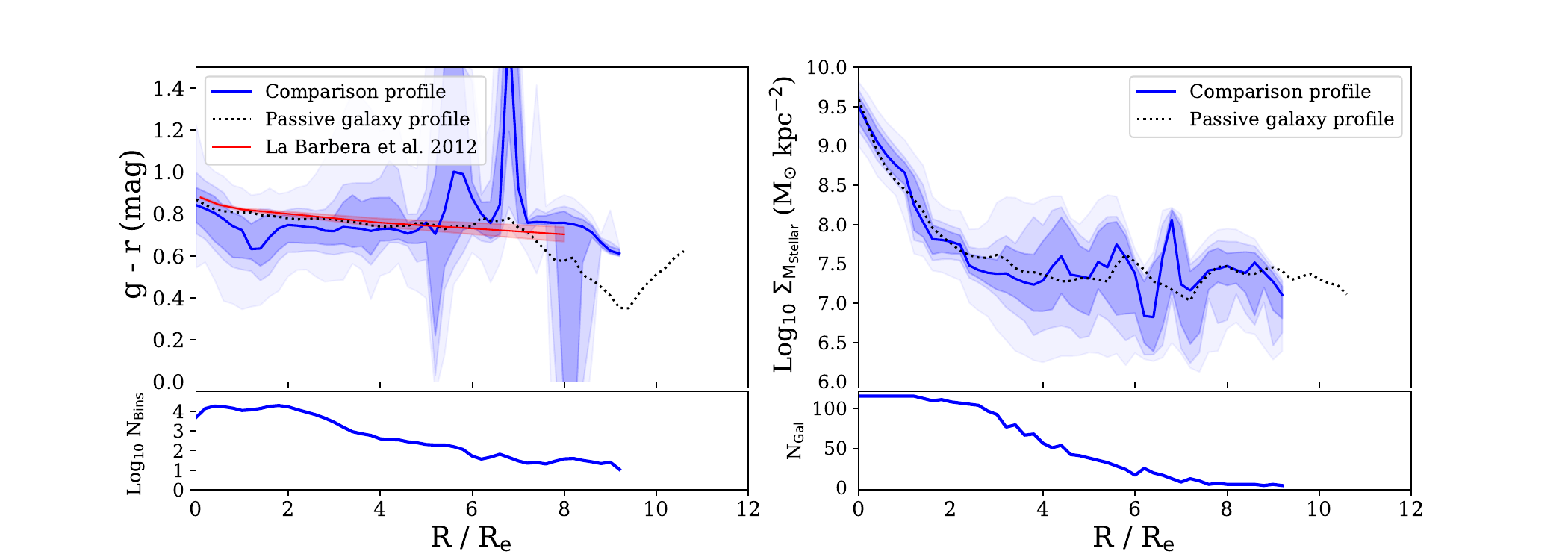}
	\caption{The top left hand panel displays the stacked radial profile (in bins of 0.1 R/R$_\mathrm{e}$) of {\it g} - {\it r} colour for the 118 younger ($t_{90} < 4$ Gyr) galaxies of the comparison sample. The solid blue line gives the median of the distribution of each bin, while the shaded areas represent the 5$^{\mathrm{th}}$, 16$^{\mathrm{th}}$, 25$^{\mathrm{th}}$, 75$^{\mathrm{th}}$, 84$^{\mathrm{th}}$ and 95$^{\mathrm{th}}$ percentiles of each stacked bin. For comparison, the results from the study of \protect{\citet{2012MNRAS.426.2300L}} are given by the solid red line and shaded areas and the median of the passive galaxy sample is given by the black dotted line. The top right hand panel shows the stacked radial profile (also in bins of 0.1 R/R$_\mathrm{e}$) of the stellar mass surface density with corresponding percentiles, normalised by the integrated stellar mass within 1 R$_\mathrm{e}$. The median of the passive galaxy sample is given by the black dotted line. The bottom left and bottom right panels show the log of the number of Voronoi bins per bin in radius and the number of galaxies contributing to the profile within any one bin in radius respectively.}
	\label{fig:comp_radial}
\end{figure*}

The top left hand panel of Figure~\ref{fig:gr_radial} shows the {\it g} - {\it r} colour profile of our passive galaxy sample. The black solid line shows the median value of the distribution in each bin of R/R$_\mathrm{e}$, the dark grey shading shows the 25$^{\mathrm{th}}$ and 75$^{\mathrm{th}}$ percentiles, the mid grey the 16$^{\mathrm{th}}$ and 84$^{\mathrm{th}}$ percentiles and the light grey the 5$^{\mathrm{th}}$ and 95$^{\mathrm{th}}$ percentiles. We see a slight decrease of the order of 0.1 dex in the {\it g} - {\it r} colour in the inner 2 R$_\mathrm{e}$, with a very narrow distribution. As we go beyond this radius towards 6 R$_\mathrm{e}$, the {\it g} - {\it r} colour still declines, becoming bluer by another 0.1 dex, however the distribution experiences much greater range (up to $\sim$ 0.5 - 1 magnitude). Beyond 8 R$_\mathrm{e}$, the stellar material is bluer, reaching a median {\it g} - {\it r} colour of 0.4 mag. On closer inspection of the maps of the galaxies predominantly contributing at these radii, this material is dominated by incoming streams and from satellite galaxies in the process of minor mergers, however we note that there are relatively few bins, meaning we have to be careful about how we interpret these results. 

We also observe a similar gradual decline of $\sim$ 0.1 mag in various other colour profiles such as ({\it g} - {\it i}, {\it r} - {\it i}, {\it i} - {\it z}) within 6 R$_\mathrm{e}$, however beyond this the  {\it r} - {\it i} and {\it i} - {\it z} colour profiles continue to gradually decline rather than dropping $\sim$ 0.4 mag. We postulate that this could be due to the amount of bins at these radii, which is much lower and hence susceptible to stochastic fluctuations at these radii. 

\begin{figure*}
	\centering
	\includegraphics[width=1.7\columnwidth]{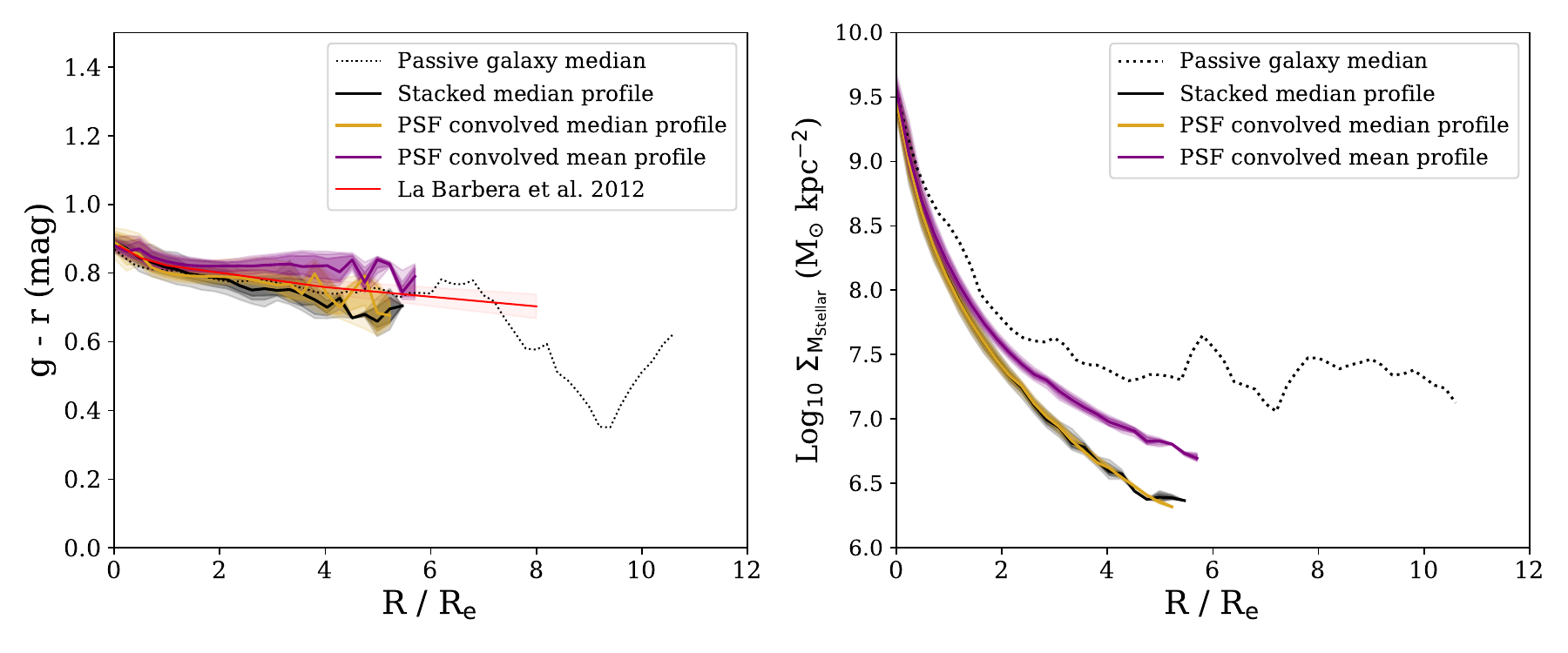}
	\caption{Left hand panel: The {\it g} - {\it r} colour profiles for our median stacked galaxy image as computed in Section~\ref{res:stacked} (solid black line), our median stacked galaxy image convolved with the constructed PSF (solid gold line), and our mean stacked galaxy image convolved with the constructed PSF (purple line). Each respective solid line indicates the median of the profile, with the various shadings representing the 5$^{\mathrm{th}}$, 16$^{\mathrm{th}}$, 25$^{\mathrm{th}}$, 75$^{\mathrm{th}}$, 84$^{\mathrm{th}}$ and 95$^{\mathrm{th}}$ percentiles of each stacked bin. For comparison the results of \protect{\citet{2012MNRAS.426.2300L}} are indicated by the solid red line and shading and the dotted black line represents the median profile from our passive galaxies in Figure~\ref{fig:gr_radial}. Right hand panel: The stellar mass surface density profile for our median stacked galaxy image as computed in Section~\ref{res:stacked} (black line) with respective percentiles, the median stacked galaxy convolved with the constructed, extended PSF (golden line), and the mean stacked galaxy convolved with the constructed PSF (purple line). For comparison the observed profile in Figure~\ref{fig:gr_radial} is represented by the black dotted line. We see that the colour profiles yielded by all stacking procedures are very similar to each other and previous results (within 0.1 mag) and that the median stacking washes out low surface brightness features, the constructed PSF has a minimal effect on the profiles and that the when we mean stack and PSF convolve our galaxy sample, the signal of the low surface brightness features is dampened.}
	\label{fig:stacked_radial}
\end{figure*}

We compare these colour profiles to a previous study carried out by \citet{2012MNRAS.426.2300L}. In their work they investigate the colour profiles of 674 massive early type galaxies, with stellar masses between 10$^{10.5}$ and 10$^{11.8}$ M$_\odot$, and contained in the SDSS and UKIRT Infrared Deep Sky Survey (UKIDSS) out to 8 R$_\mathrm{e}$. Their results are represented by the solid red line for the median and the red shading indicating the 16$^{\mathrm{th}}$ and 84$^{\mathrm{th}}$ percentiles. We find good agreement between our results and those from \citet{2012MNRAS.426.2300L} as the values are well within the scatter of our distribution at almost all radii. 

The top right hand panel of Figure~\ref{fig:gr_radial} shows the stellar mass surface density profile as yielded by the estimates of stellar mass from \textsc{cigale}. Once again the median of each distribution in a bin of 0.1 R/R$_{\mathrm{e}}$ is represented by the black solid line with the 5$^{\mathrm{th}}$, 16$^{\mathrm{th}}$, 25$^{\mathrm{th}}$, 75$^{\mathrm{th}}$, 84$^{\mathrm{th}}$ and 95$^{\mathrm{th}}$ percentiles given by the various shading levels. We see a fairly smoothly declining profile within $\sim 3$ R$_{\mathrm{e}}$, as expected, but beyond this radius we observe a flattening of the average stellar mass density profile, with some bumps. This behaviour occurs not only in the median values of the distribution but also in various percentiles of the distribution. This indicates that significant stellar material is present in the observed tidal features we see in the imaging. 

The bottom left and bottom right panels of Figure~\ref{fig:gr_radial} show the log of the number of Voronoi bins per bin in radius and the number of galaxies contributing to the profile within any one bin in radius respectively. We note that, as may be expected with our technique, the number of Voronoi bins and the number of galaxies contributing to each bin in R/R$_{\mathrm{e}}$ decreases significantly in the outermost radii.

Figure~\ref{fig:comp_radial} shows the same radial profiles for our comparison galaxy sample, with the median (in bins of 0.1 R/R$_\mathrm{e}$) represented by the solid blue line and the shaded regions indicating the 5$^{\mathrm{th}}$, 16$^{\mathrm{th}}$, 25$^{\mathrm{th}}$, 75$^{\mathrm{th}}$, 84$^{\mathrm{th}}$ and 95$^{\mathrm{th}}$ percentiles, the same as for our passive galaxy sample in Figure~\ref{fig:gr_radial}. The median {\it g} - {\it r} colour for our passive galaxy sample is given by the dotted black line for comparison. We see slightly bluer average colours of the order of $\sim$ 0.1 mag in the central regions of our galaxies up until $\sim$ 5 R$_\mathrm{e}$. Beyond this point, much of the material is redder than our passive galaxy sample or the measured profiles of \citet[][given by the solid red line]{2012MNRAS.426.2300L}. We note, however, that the profile beyond this point relies on less Voronoi bins, which introduces large variations. These spikes are predominantly driven by streams of accreted material and minor merger remnants, similar to the passive galaxy sample. 

The stellar mass density profile of our comparison sample (median represented by the solid blue line with shaded regions showing the various percentiles of the distribution) shows similar behaviour to our passive galaxy sample (median represented by the dotted black line), whereby we observe a steeply declining profile in the inner parts of the galaxy before a flattening of the profile at larger radii with some bumps. Similar to Figure~\ref{fig:gr_radial} for the passive sample, the bottom right and bottom left panels of Figure~\ref{fig:comp_radial} show the log of the number of Voronoi bins per bin in radius and the number of galaxies contributing to the profile within any one bin in radius respectively.

\subsection{Profiles yielded from a stacked image}
\label{res:stacked}

Some previous works have carried out similar studies of the growth processes of massive galaxies including the stellar mass content of their outskirts such as that of \citet{2010ApJ...709.1018V}. In their study, they took imaging data from a number of surveys such as SDSS and their own NEWFIRM Medium Band Survey and stacked the galaxy images in bins of redshift, normalised and constructed radial profiles of the stellar mass density. Their profiles showed an increase in the average stellar mass density in the outskirts of their sample with decreasing redshift, which drives the increase in effective radius and thereby the measured size of their galaxy sample. These profiles, however, are smooth in contrast to our passive galaxy and comparison galaxy profiles. This is because averaging, either through stacking or isophotal analysis washes out or dampens the signal of specific merger features which generally have an asymetric orientation on the sky compared to the host galaxy. To check that this hypothesis holds, we employed similar processes, stacking our galaxy images and running source detection, Voronoi binning, and SED fitting, to see if we replicate similar trends to \citet{2010ApJ...709.1018V}.

Additionally, as stated earlier, the wings of the PSF may also have an effect on surface brightness profiles, and thereby on the colour and stellar mass surface density profiles. To investigate the effect this may have on our results, and whether the PSF wings may account for the flattening in our profiles in Section~\ref{res:rad_gradients}, we constructed our own, extended PSFs to be convolved with our stacked galaxy image described above. We carry out this qualitative test in order to see how much of an effect the PSF wings may have on the stacked galaxy profiles. A detailed analysis of the PSF at large radii is beyond the scope of this project.

We selected 15 stars in the area of the sky covered by our galaxy sample that are over-saturated in all five of the HSC survey filters, in order to well-sample the wings of the extended PSF. For each filter, we also masked out fore- and back-ground sources and computed their median image. We corrected the PSF median image for residual sky background by fitting a 2D surface through the sky pixels and then subtracting it. To deal with the over-saturated centre, we replaced the saturated core with a scaled, unsaturated PSF. This was done by selecting 20 field stars which are unsaturated in all of the HSC bands. We treated their images in the same way as the the selected saturated stars, isolating foreground and background objects, median stacking, and modelling and subtracting the sky. The final composite PSF in each filter was then normalised. Such a constructed, extended PSF reaches out to a radius beyond 100 arcsec, comparable with the maximum extension of the galaxies in our sample.

We then took the galaxy images with the extra sky background subtraction and convolved to the worst seeing (as described in Section~\ref{OM:detect_binning} in each band) and magnified each image to the median redshift of 0.086 using \textsc{iraf}\footnote{http://ast.noao.edu/data/software} in order to bring each galaxy onto the same physical spatial scale while conserving the flux (i.e. the sum of the total fluxes in all of the old and new pixels is equal). We normalised each flux value by the integrated stellar mass as in \citet{2010ApJ...709.1018V}, corrected for redshift dimming and then stacked all images in each band. We then stack all images, taking both the mean, in line with previous observational studies \citep[e.g.][]{2010ApJ...709.1018V,2014MNRAS.443.1433D}, and the median, to show a galaxy without any signal from merger or accretion features, for all stacked pixels. We then applied the same Voronoi binning procedure as before to the median stacked galaxy, measured the {\it g} and {\it r} colours, and fitted the photometry for each Voronoi bin with an SED to yield stellar mass estimates for the stellar mass surface density profiles. We then constructed similar radial profiles as the previous section. We also multiplied the effective radii of each galaxy by the same magnification factors yielding a median angular effective radius of 3.4 arcsec, which we use in order to compare our stacked galaxy profile to our radial profiles in Section~\ref{res:rad_gradients}. The results for the median stacked galaxy are shown in Figure~\ref{fig:stacked_radial}, with the solid black line representing the median with the accompanying grey shaded areas representing the same percentile limits as Figures~\ref{fig:gr_radial} and \ref{fig:comp_radial}.

We then took both the median and mean stacked galaxy images in each band and fitted each with a two component profile containing a bulge and disk contribution in two dimensions, masking stars captured in the mean stacked galaxy to avoid biasing the fits. We then convolved each fitted profile with the constructed PSFs in each band, to produce images of a model median or mean galaxu convolved with the observed PSFs. We ran the same process of Voronoi binning and SED fitting using \textsc{cigale} in order to estimate stellar mass surface densities, and subsequently used these to construct the profiles seen in Figure~\ref{fig:stacked_radial}, referred to as the PSF convolved median (gold) and PSF convolved mean (purple) galaxy with the accompanying shaded areas as in previous figures.

\begin{figure*}
	\centering
	\includegraphics[width=1.8\columnwidth]{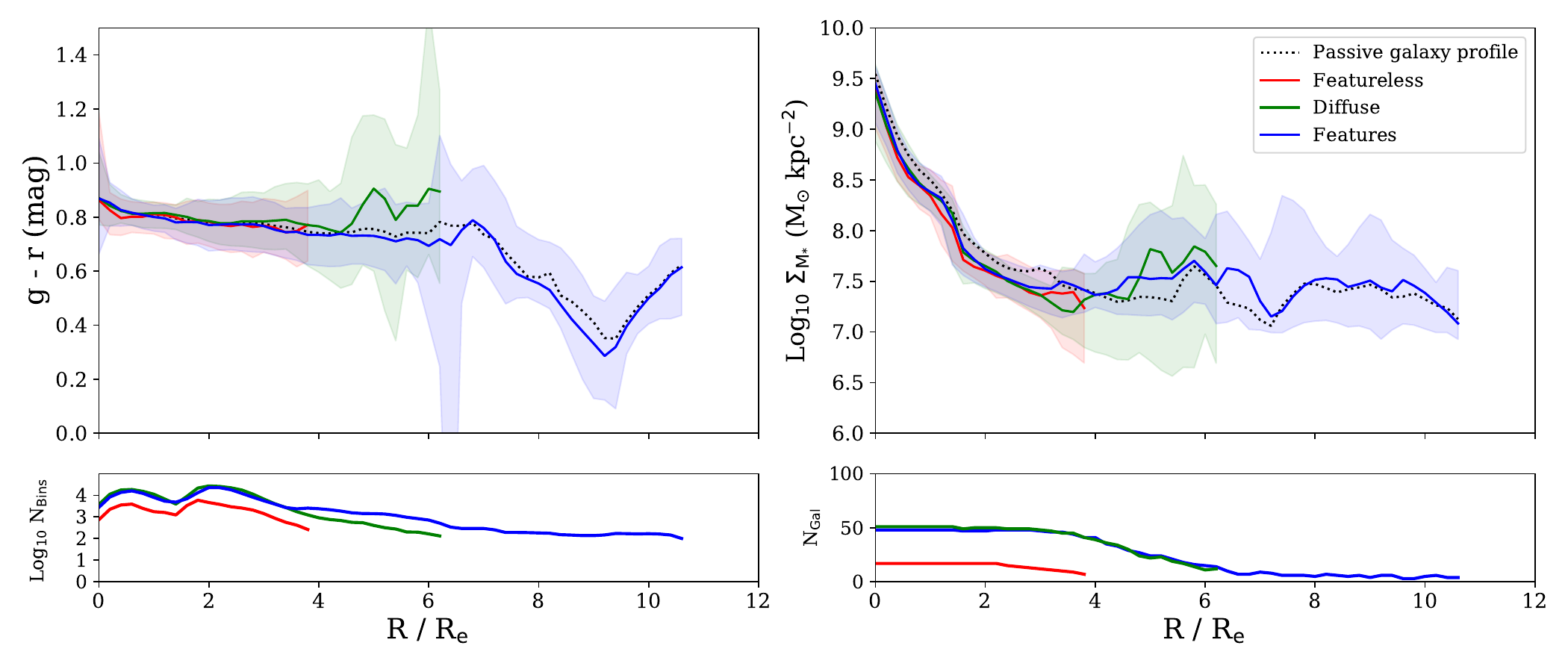}
	\caption{Top left hand panel: The {\it g} - {\it r} colour profiles of those galaxies that display signs of merger activity (features, median represented by the blue solid line and 16$^{\mathrm{th}}$ and 84$^{\mathrm{th}}$ percentiles represented by the blue shaded region), those that do not display clear signs of merger activity but a diffuse stellar halo (diffuse, median represented by the green solid line and 16$^{\mathrm{th}}$ and 84$^{\mathrm{th}}$ percentiles represented by the green shaded region) and those that display no merger activity or diffuse halo (featureless, median represented by the red solid line and 16$^{\mathrm{th}}$ and 84$^{\mathrm{th}}$ percentiles represented by the red shaded region). Top right hand panel: The stellar mass density profile for each of the three aforementioned sub-samples. The bottom left and bottom right panels show the log of the number of Voronoi bins per bin in radius and the number of galaxies contributing to the profile within any one bin in radius for each of the sub-samples respectively.}
	\label{fig:split_radial}
\end{figure*}

We observe little difference in the {\it g} - {\it r} colour profiles in the median stacked galaxy profile (solid black line and accompanying grey shaded areas showing the various percentiles) and when convolved with the constructed PSF (median given by the golden line with corresponding shaded areas showing the various percentiles), with differences less than 0.1 mag at almost all radii. The stacked galaxy stellar mass density profile appears as a smooth decline out to $\sim$ 4 R$_{\mathrm{e}}$, similar to previous works, with some minor deviations beyond these radii. When we compare, we see that the stellar mass density profile of the stacked galaxy convolved with our constructed, extended PSF follows the stellar mass density profile of the stacked galaxy up to $\sim$ 3 R$_\mathrm{e}$. Slight deviations are present beyond this, however these are less than 0.1 dex and less than the uncertainty on our measurements. These differences are much smaller than the differences between either of these two profiles and our passive galaxy median profile, with differences over an order of magnitude displayed between the stacked galaxy profile/PSF convolved profile and the passive galaxy median profile. In comparison, the mean stacked profiles convolved with the constructed PSF are represented by the purple line and respective shaded areas. We see similar {\it g} - {\it r} colour profiles in the central regions, with some differences of 0.1 mag in the outskirts. The stellar mass surface density profile of this galaxy, however, although similar in the central regions, displays an excess compared to the median profiles at larger radii. This difference is of the order of 0.5 dex, and is likely due to the added signal from the merger features increasing the average, with minor contamination from foreground and background sources not completely accounted for in the masking process. This excess, however, is not as great as in our passive galaxy sample profile, which is likely due to the averaging including the background, as mentioned above. We therefore come to the conclusion that although PSF effects may contribute minorly to the stellar mass density seen in the outskirts of galaxies using our techniques from Section~\ref{OM:detect_binning}, they are not the main driver of the excesses, meaning that the bumps seen in the stellar mass density profiles are driven by the tidal features we specifically detect using our methods described in Section~\ref{OM:detect_binning}.

Using the median redshift and effective radius of our sample, we can calculate an approximate median physical effective radius of our stacked profile of 5.9 kpc. We can therefore approximately compare our stellar mass surface density profile of the median and mean stacked, PSF corrected (golden and purple respectively) profiles to that of \citet{2010ApJ...709.1018V}. We choose physical radii of 10, 20 and 30 kpc: \citet{2010ApJ...709.1018V} find stellar mass densities of $\sim$ 10$^{8}$, 10$^{7.2}$ and $\sim$ 10$^{7}$ M$_\odot$ kpc$^{-2}$ at 10, 20 and 30 kpc respectively, that compares to our results of $\sim$ 10$^{7.9}$, 10$^{7.1}$ and 10$^{6.9}$ M$_\odot$ kpc$^{-2}$ and $\sim$ 10$^{7.6}$, 10$^{6.7}$ and 10$^{6.3}$ M$_\odot$ kpc$^{-2}$ for the mean and median profiles at 10, 20 and 30 kpc respectively. We therefore see a very similar behaviour in our profiles compared to \citet{2010ApJ...709.1018V}, with less than 0.1 dex differences in our mean profiles, and slight, but expected, dearth in the median. These slight differences may be due to different data sets (i.e. SDSS and NEWFIRM imaging data compared to HSC) or the different methods used to treat the data, such as that the study of \citet{2010ApJ...709.1018V} assume a \citet{2001MNRAS.322..231K}, whereas we assume a \citet{2003PASP..115..763C} in our SED fitting.

\subsection{Profiles of sub-samples: the driver of the excesses}
\label{res:splits}

In order to further investigate the nature of the bumps in the passive galaxy stellar mass density radial profiles, we split the sample by the morphological classifications as described in Section~\ref{OM:classification}, namely the features, the diffuse and the featureless sub-samples outlined in Section~\ref{OM:classification}. We then plot the same radial profiles as in the two previous sections. Various examples of individual profiles can be found in Figures~\ref{fig:features_profiles}, ~\ref{fig:diffuse_profiles} and ~\ref{fig:featureless_profiles} for the features, diffuse and featureless sub-samples respectively.

We see in Figure~\ref{fig:split_radial} that all sub-samples have extremely similar {\it g} - {\it r} colour profiles inside $\sim$ 4 R$_{\mathrm{e}}$, decreasing from $\sim$ 0.9 mag in the centre to $\sim$ 0.8 mag by 4 R$_{\mathrm{e}}$. The featureless galaxies (solid red line for the median, 16$^{\mathrm{th}}$ and 84$^{\mathrm{th}}$ percentiles are shaded) do not extend out beyond this radius. The diffuse galaxies (solid green line for the median, 16$^{\mathrm{th}}$ and 84$^{\mathrm{th}}$ percentiles are shaded) increase in {\it g} - {\it r} colour by 0.1 mag to $\sim$ 0.9 mag beyond this radius. The features sub-sample (solid blue line for the median, 16$^{\mathrm{th}}$ and 84$^{\mathrm{th}}$ percentiles are shaded) dominates the sample number counts at high radii, and follows a similar profile to the overall passive galaxy sample (dotted black line for the median). 

We see that the stellar mass surface density profiles for all sub-samples decrease in the central regions, until $\sim$ 3 R$_{\mathrm{e}}$. For the featureless sub-sample, this trend appears to generally decrease, but cuts off at 4 R$_{\mathrm{e}}$, meaning they do not contribute to the flattening of the stellar mass surface density profiles we observe. For the diffuse sub-sample, we see a flattening off of the profile beyond 3 R$_{\mathrm{e}}$ and then a significant decrease, possibly due to a truncation, in the stellar mass surface density at 6 R$_{\mathrm{e}}$. The features sub-sample (solid blue line for the median, 16$^{\mathrm{th}}$ and 84$^{\mathrm{th}}$ percentiles are shaded) shows very similar behaviour to the diffuse sub-sample, initially flattening off at 3 R$_{\mathrm{e}}$. However, beyond 6 R$_{\mathrm{e}}$, this sub-sample dominates the excess material we detect, remaining flat with some bumps. On visual inspection, these bins are dominated by galaxies that are undergoing minor mergers or have stellar streams. This shows that the stellar material in the outskirts of our passive galaxy sample is dominated by the tidal features seen in these galaxies that are indicative of recent galaxy interactions and/or mergers. 

\subsection{Exploration of biases in the stellar mass estimates}
\label{disc:biases}

One source of bias in our results may come from the SED fitting we apply to the imaging data. In order to check that our stellar mass estimates are consistent with previous work, we compare total stellar mass estimates from our SED fitting using \textsc{cigale} on the HSC imaging data to the stellar mass estimates from the parent sample in \citet{2020MNRAS.497.4262J}, which are derived from \citet{2016ApJ...824...45P} \citep[see Section~\ref{sec:Data} or ][for more detail]{2012MNRAS.421.2002P, 2016ApJ...824...45P}. We measure the fluxes within the same apertures sizes (SDSS Petrosian radius) from our HSC imaging data as those used in \citet{2016ApJ...824...45P} for consistency.  

\begin{figure}
	\centering
	\includegraphics[width=1.0\columnwidth]{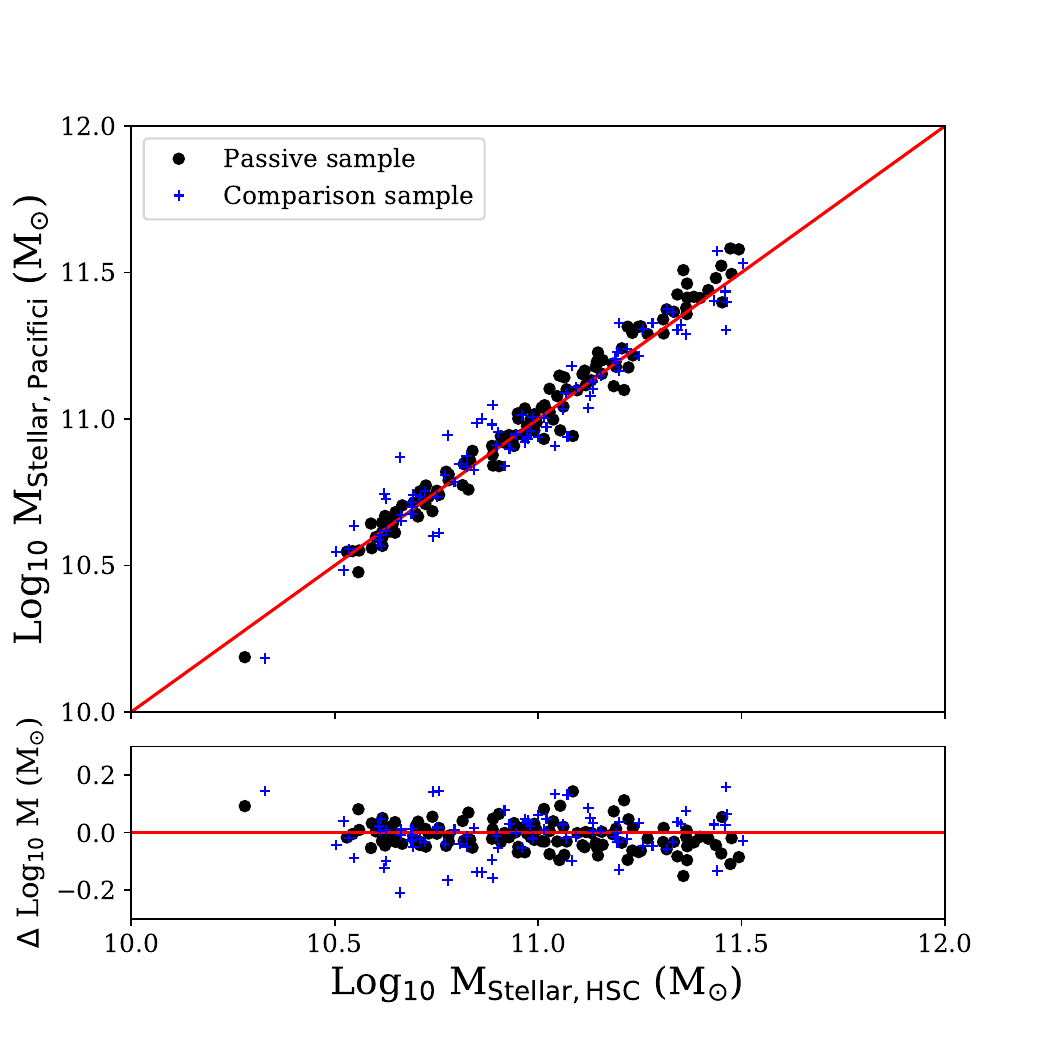}
	\caption{Top panel: A comparison of the stellar mass estimates from the study of \protect\citet{2016ApJ...824...45P} with the estimates yielded from \textsc{cigale} for the central regions of the HSC data. Bottom panel: The difference in log space between the two different methods.}
	\label{fig:sm_comp}
\end{figure}

The results of our comparison of the stellar masses are shown in Figure~\ref{fig:sm_comp}. In the top panel, we see a tight correlation between the estimates from the HSC data computed by \textsc{cigale} and the routine from \citet{2016ApJ...824...45P}. The bottom panel of Figure~\ref{fig:sm_comp} shows that there are no obvious systematic trends or outliers in both our passive galaxy sample and the comparison sample, with maximum differences in the estimates of $\sim$ 0.2 dex (within the observational uncertainties). This confirms that our data and SED fitting routine are consistent with previous estimates, and therefore should have a good level of reliability when applied to the outskirts of our galaxy sample.  

The star formation history used in generating models for SED fitting routines is also a potential source of bias. Recent studies such as that of \citet{2020ApJ...893..111L} have shown that using different star formation histories, including non-parametric ones, can cause differences of 0.1 - 1 dex in the stellar mass estimates. In order to check this potential source of bias we also ran our SED fitting with \textsc{cigale} assuming a delayed star formation history on all Voronoi bins in the passive galaxy sample, including those in the outskirts of the galaxies. The estimates of stellar mass yielded had a median difference of 0.1 dex with a maximum difference of 0.3 dex, generally within the estimated uncertainties of the stellar mass. This means that the radial profiles did not change significantly. This combined with the agreement between our estimates for the central regions of the galaxy samples and those from \citet{2016ApJ...824...45P}, which use non-parametric star formation histories, indicates our SED fitting does not significantly bias our results.  

Redshift biases may also be present in our morphological classification. At larger distances or higher redshifts, the sensitivity level of the survey means that we do not probe to as deep in surface brightness, meaning that some low surface brightness features may be missed and a galaxy classified as displaying no merger activity when there may be activity below the sensitivity level of the survey. To investigate possible biases we plot the redshift distributions of each sub-sample in redshift space and compare them in the bottom right panels of Figure~\ref{fig:sample}.

We see that both in the passive galaxy sample and the comparison sample there is significant overlap in redshift between the three sub-samples and that any differences in the average redshift are minimal ($\Delta z <$ 0.002). We also searched imaging data from the HSC ultra deep survey fields for galaxies in both samples which we classified as featureless. Although only 6 galaxies were available in these fields, none displayed any low surface brightness features when using imaging two orders of magnitude deeper. We argue that this shows that we are not significantly redshift or sensitivity biased in our morphological classification.  

\section{Discussion}
\label{sec:Discussion}

\subsection{Comparison of the colours profiles to satellites}
\label{disc:satellites}

One way to analyse the nature of the material we observe in the outskirts of these galaxies is to investigate the colours of various populations of satellite galaxies in comparison to the outskirts themselves.  

\begin{figure}
	\centering
	\includegraphics[width=1.0\columnwidth]{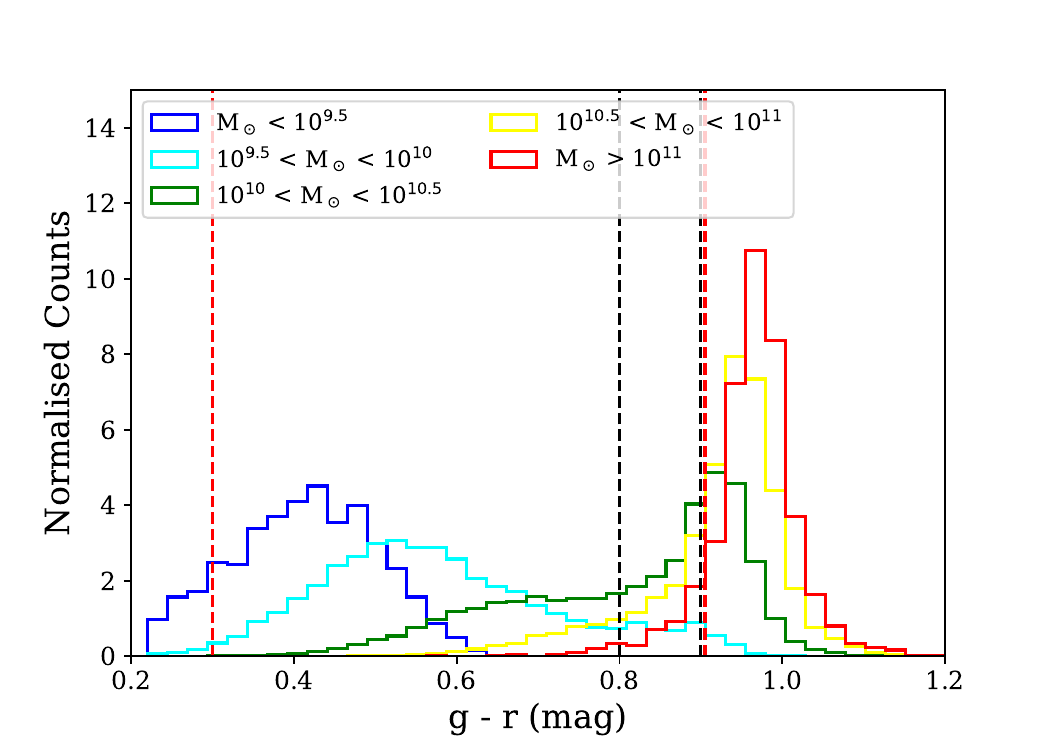}
	\caption{The distribution in {\it g} - {\it r} colour of satellites taken from SDSS catalogues in bins of 0.5 dex in stellar mass. The dashed black and red lines mark the approximate colour limits of the bulk of stellar material in the centre (< 2 R$_{\mathrm{e}}$) and outskirts (> 2 R$_{\mathrm{e}}$) of our passive galaxy radial profiles respectively.}
	\label{fig:sat_colours}
\end{figure}

To do this, we take the SDSS catalogues of \citet{2017MNRAS.470.2982L} and select only satellite galaxies, defined as those galaxies in a halo that are not designated as the central galaxy of that halo, across the same redshift range as our passive galaxy sample (0.05 < $z$ < 0.1). We then calculate the {\it g} - {\it r} colour of each satellite from the SDSS photometry and bin in intervals of 0.5 dex in stellar mass. The distributions can be seen in Figure~\ref{fig:sat_colours}. We see the well known trend whereby more massive galaxies tend to be on average redder in their {\it g} - {\it r} colour.

The median {\it g} - {\it r} colour profile in the inner parts of our passive galaxy sample decreases from 0.9 to 0.8 mag between 0 and 6 R$_{\mathrm{e}}$ (indicated by the black dashed lines in Figure~\ref{fig:sat_colours}), with an average scatter of $\sim$ 0.1 mag. As we increase the radius beyond 6 R$_{\mathrm{e}}$, we observe a range of median {\it g} - {\it r} colours, from $\sim$ 0.4 to 0.8 mag, with increased scatter in the {\it g} - {\it r} colour range $\sim$ 0.5 mag. The approximate boundaries of the profiles in these regions (> 6 R$_{\mathrm{e}}$) are indicated by the red dashed line in Figure~\ref{fig:sat_colours}. The comparison sample displays similar behaviour in the nature of the stellar material inside of 6 R$_{\mathrm{e}}$, with median {\it g} - {\it r} colours ranging between 0.9 and 0.7 mag, however with greater scatter ($\sim$ 0.3 mag). The profiles in the outer regions exhibit a greater range of {\it g} - {\it r} colours between 0.2 and 1.2 mag, probably because of stochatisticity due to the small sample and of the merger activity. 

When we compare to the distribution of SDSS satellites in Figure~\ref{fig:sat_colours}, we see that those satellite galaxies ranging in stellar mass M$_{\mathrm{Stellar}}$ < 10$^{10.5}$ M$_\odot$ have the most similar colours to the low surface brightness material. Some of the redder material may be linked to higher mass satellites, however this is not the majority of the material beyond 6 R$_{\mathrm{e}}$. If we account for a possible contribution from redder, in-situ stars with similar colours to the profiles within 6 R$_{\mathrm{e}}$, we would expect the accreted material to have even bluer average colours. This would shift potential accreted satellite galaxies to even bluer colours and, on average, smaller stellar masses.

\subsection{Interpretation of the level of activity and radial profile excess}
\label{disc:interpretation}

\begin{table*}
	\centering
	\caption{The total stellar mass in bins of 1 R/R$_{\mathrm{e}}$ }
	\label{tab:percentages}
	\begin{tabular}{lcccccccr}
		\hline
		Radius & Passive (features) & & Passive (diffuse) & & Comparison (features) & & Comparison (diffuse) & \\
		(R$_{\mathrm{e}}$) & Log$_{10}$ M (M$_\odot$) & Percent  & Log$_{10}$ M (M$_\odot$) & Percent & Log$_{10}$ M (M$_\odot$) & Percent  & Log$_{10}$ M (M$_\odot$) & Percent \\
		\hline
		Total & 11.41  & - & 11.42 & - & 11.32 & - & 11.39 & - \\
		<1 R$_{\mathrm{e}}$ & 11.30  & 77.69\% & 11.32 & 79.23\% & 11.22 & 78.20\% & 11.29 & 80.16\% \\
		1 - 2 R$_{\mathrm{e}}$ & 10.52  & 12.77\% & 10.54 & 13.10\% & 10.49 & 14.50\% & 10.51 & 13.19\% \\
		2 - 3 R$_{\mathrm{e}}$ & 10.18  & 5.87\% & 10.12 & 4.97\% & 9.99 & 4.64\% & 10.04 & 4.53\% \\
		3 - 4 R$_{\mathrm{e}}$ & 9.72 & 2.07\% & 9.70 & 1.90\% & 9.51 & 1.55\% & 9.48 & 1.26\% \\
		4 - 5 R$_{\mathrm{e}}$ & 9.30 & 0.78\% & 9.12 & 0.49\% & 9.09 & 0.58\% & 9.15 & 0.58\% \\
		5 - 6 R$_{\mathrm{e}}$ & 9.04 & 0.43\% & 8.83 & 0.25\% & 8.81 & 0.31\% & 8.56 & 0.15\% \\
		6 - 7 R$_{\mathrm{e}}$ & 8.55 & 0.14\% & 8.25 & 0.06\% & 8.41 & 0.12\% & 8.52 & 0.13\% \\
		7 - 8 R$_{\mathrm{e}}$ & 8.49 & 0.12\% & - & - & 8.05 & 0.05\% & - & - \\
		> 8 R$_{\mathrm{e}}$ & 8.53 & 0.13\% & - & - & 8.05 & 0.05\% & - & -\\
		\hline
	\end{tabular}
\end{table*}

In Section~\ref{res:rad_gradients}, we showed tidal features due to accreted material can be quantified in colour and stellar mass surface density. These tidal features are reflected as a flattening in the outskirts of stellar mass surface density profiles ($\gtrsim 3$ R$_{\mathrm{e}}$) when trying to specifically detect and quantify these features. We see from Sections~\ref{res:stacked} and \ref{res:splits} that PSF effects, although shifting some of the stellar mass surface density from the centres of the radial profiles to the outskirts, cannot account for the levels of stellar mass surface density at radii of $\gtrsim 3$ R$_{\mathrm{e}}$. We have also seen that the stellar mass surface densities are driven predominantly by galaxies that show features, hence signs of merger activity, with some minor contributions by galaxies which display a diffuse stellar halo. We also showed that the material in the outskirts of our galaxy samples has similar {\it g} - {\it r} colours to satellite galaxies ranging in stellar mass M$_{\mathrm{Stellar}}$ < 10$^{10.5}$ M$_\odot$. 

In order to further link the stellar mass excesses to possible satellite galaxy accretion, we quantify the total stellar mass of this material in both our passive and comparison galaxy samples, as a large percentage of the stellar mass contained in the outskirts may indicate different origins of this material. We firstly isolate the sub-samples of galaxies that display features in both the passive and comparison galaxy samples. We then calculate the average stellar mass per galaxy contained in a bin of 1 R/R$_{\mathrm{e}}$ from the estimates of stellar mass yielded by the SED fitting process. We then work out the percentage stellar mass that each bin in 1 R/R$_{\mathrm{e}}$ contributes to the total stellar mass contained on average in one galaxy. These results can be seen in Table~\ref{tab:percentages}.

We firstly see that for both our passive and comparison galaxy samples, those galaxies that display features contribute more stellar mass in their outskirts. The integrated stellar mass in these regions, however, is relatively small; as outside of 2 R$_{\mathrm{e}}$ for the passive galaxy sample this totals a stellar mass of 10$^{10.4}$ M$_\odot$ (equating to $\sim$ 9 per cent) for the galaxies which display features and 10$^{10.3}$ M$_\odot$ (equating to $\sim$ 8 per cent) for our diffuse sub-sample. For the comparison sample, the stellar mass contained outside of 2 R$_{\mathrm{e}}$ is 10$^{10.2}$ M$_\odot$ (or $\sim$ 7 per cent) both for galaxies that exhibit features and those that have a diffuse halo. 

Taking the median integrated stellar mass estimates at all radii (denoted as `Total') from Table~\ref{tab:percentages} of $\sim$ 10$^{11.3 - 11.4}$ M$_\odot$ and convolving them with the standard definition of a minor merger as 1:4 < $\mu$ < 1:100 \citep[as in][]{2015MNRAS.449..528H}, we get an average stellar mass of the accreted satellites of $\sim$ 10$^{9.3 - 10.8}$ M$_\odot$. Comparing this figure to the integrated stellar mass  in the outskirts (outside of 2 R$_{\mathrm{e}}$) of 10$^{10.2 - 10.4}$ M$_\odot$, we see that a minor merger is capable of delivering the amount of stellar mass required to explain the stellar material found in the outskirts.

The amount of mass required to be accreted is, however, also likely to be smaller than this estimate, as not all stars at these radii are ex-situ. \citet{2020MNRAS.497...81D} find that some in-situ stars exist at these radii using simulations, whereby the ex-situ fraction of stars in galaxies of stellar mass 10$^{10}$ M$_\odot$ - 10$^{11}$ M$_\odot$ is $\sim$ 25 per cent at 2 R$_{\mathrm{e}}$, up to 40 per cent at 4 R$_{\mathrm{e}}$. This is increased to 65 and 70 percent, respectively, for galaxies of stellar masses of 10$^{11}$ M$_\odot$ - 10$^{12}$ M$_\odot$. This significantly lowers the required mass to be delivered by the satellite galaxies by 30 - 60 percent.

To summarise: The average stellar mass per galaxy contained in the outskirts of our sample, mainly in the form of tidal and merger features is of the order of 10$^{10.2 - 10.4}$ M$_\odot$. The ex-situ mass contained here is likely to be slightly lower as not all stars at these radii (> 2 R$_{\mathrm{e)}}$) are ex-situ. The colour of this material is similar to the average SDSS satellite population with stellar masses M$_{\mathrm{Stellar}}$ < 10$^{10.5}$ M$_\odot$. This leads to the plausible scenario that this material was accreted from the surrounding satellite population. These results, combined with the widespread merger activity we see ($\sim$ 40 per cent in both samples) quantitatively strengthen the scenario where minor mergers drive the size growth of central galaxies observed in previous studies \citep[e.g.][]{2010ApJ...709.1018V, 2014ApJ...788...28V}.

\subsection{Comparison to previous observational studies}
\label{disc:comparison}

One factor to keep in mind for in the comparison of low surface brightness features to merger events is the survival time of various merger features. A recent study by \citet{2019A&A...632A.122M} used {\it N}-body simulations to compare the formation of merger features or tidal activity to observational data and calculated the average survival time of these types of features. They estimated a survival time of $\sim$ 2 Gyr for tidal tails, $\sim$ 3 Gyr for streams and $\sim$ 4 Gyr for shells, however, do not state how this relates to the surface brightness these features are detectable at. As the study uses surface brightness limits of 29 and 33 mag arcsec$^{-2}$, we assume these times to be longer than what we would expect with our imaging, which only goes down to $\mu_{r\mathrm{-band}}$ $\sim$ 28 mag arcsec$^{-2}$. As our passive galaxies are selected to have assembled 90 per cent of their stellar mass over 6 Gyr ago, this means that the features are extremely unlikely to be due to major mergers, as a major merger with a stellar mass ratio of at least 1 to 4 would deliver more than 10 percent of the stellar mass observed today. Any activity therefore can not be older than this time, fitting the picture of these tidal features caused by minor mergers.

Turning to other works which harness simulations to investigate the stellar content in the outskirts of massive galaxies, both \citet{2013MNRAS.434.3348C} and \citet{2014MNRAS.444..237P} have investigated the stellar content of galaxy outskirts as function of the galaxy or halo mass properties. \citet{2013MNRAS.434.3348C} find a number of tidal features in their galaxy sample, with both studies finding that the fraction of accreted stars increases with galaxy stellar mass, whereby higher mass galaxies have a higher fraction of accreted stars. Both \citet{2013MNRAS.434.3348C} and \citet{2014MNRAS.444..237P} fit profiles to the stellar mass density of their sample, although we caution that both studies use a form of symmetric fitting, thereby smoothing some of the results. \citet{2013MNRAS.434.3348C} find a prominent break, whereby the profile has a significant deviation which can usually be fitted by a secondary component at radii beyond 10 kpc, at similar radii to where we observe a flattening in the profiles due to our method. \citet{2013MNRAS.434.3348C} find that this radius increases with stellar mass whereas \citet{2014MNRAS.444..237P} find a flattening of the profiles with increasing halo mass. The smoother results of \citet{2014MNRAS.444..237P} may be due to slightly different modelling or averaging methods, especially as they fit single profiles, but agree to first order. Interestingly \citet{2014MNRAS.444..237P} also find that older galaxies also have shallower slopes, likely indicating higher fractions of accreted stars. This would fit into a minor merger accretion scenario, as these older galaxies have had more time to accrete such objects. This may lend credence to our argument below that the diffuse sub-sample in our paper here may arise from already settled, accreted material.

In more recent work in simulations, \citet{2021A&A...647A..95P} explored the stellar haloes and accretion histories of a number of early type galaxies (ETGs) in the IllustrisTNG simulations. They find an increase in the accreted fraction of stars with stellar mass, which correlates with the kinematic properties of their galaxies. They also classify their sample into four different classes, dependent on whether the galaxy is dominated by in-situ stellar material, ex-situ dominated or goes from in- to ex-situ dominated from centre to outskirts or vice-versa. \citet{2021A&A...647A..95P} find the majority of their sample to be in- to ex-situ dominated, especially at intermediate stellar masses. However, this transition from in- to ex-situ dominated stellar material occurs at 6 effective radii, higher than most other studies. This may be due to variations in the techniques used to tag particles as belonging to the host galaxy or not, resulting in differences as to how particles are classified as in- or ex-situ, or sample selection, whereby \citet{2021A&A...647A..95P} select only ETGs, which could mean mass is further centrally concentrated and the effective radii smaller, pushing the transition radius to larger values.

Observations have found similar results, for example a study carried out by \citet{2016ApJ...823..123T} which imaged the galaxy UGC 00180 with the Gran Telescopio Canarias found a surface brightness profile in good agreement with the profiles found in \citet{2013MNRAS.434.3348C}. \citet{2016ApJ...823..123T} also find increasing amounts of tidal features with increasing depth of observation \citep[echoing the results of][]{2019A&A...632A.122M}. They also estimate that the stellar halo of this galaxy contributes $\sim4 \times 10^{9}$ M$_{\odot}$, or roughly 3 percent of the total stellar mass of the system, slightly lower than the 9 percent of stellar mass outside of 2 R$_\mathrm{e}$, albeit at much higher radii than 2 R$_\mathrm{e}$, meaning the difference between the two figures is likely to be much less.

Other observational studies, such as that of \citet{2014MNRAS.443.1433D}, have used much more statistically significant samples such as from SDSS to find similar results. \citet{2014MNRAS.443.1433D} implement stacking methods on the imaging to reach deeper magnitudes, using multi-component fits to model the surface brightness profiles. As with other studies, they find break radii in the profiles and an estimated fraction of accreted material varies between 2 and 70 percent. \citet{2014MNRAS.443.1433D} also find that when they can quantify the colour of the stellar halo of the galaxy, bluer colours than the main galaxy are present by between 0.5 and 1 magnitude, similar to the behaviour seen in our profiles. They also find an additional reddening of the stellar halo with increasing stellar mass.

Another example of more recent work on observational profiles is the study of \citet{2021A&A...649A.161S}, where they investigated the surface brightness profiles of three individual galaxies using deep imaging data from the VEGAS survey. They reach surface brightness levels of $\mu_{\mathrm{r}} \sim 30$ mag arcsec$^{-2}$, finding that their profiles are best fitted with a double S\'{e}rsic profile with an additional, faint exponential component at the furthest and faintest reaches of their profiles. They find that the break radii between these components is at roughly 7kpc and 15kpc respectively, similar to previous studies.

Finally, we can also compare some of our results to those found in previous studies of massive galaxies in HSC. \citet{2018MNRAS.475.3348H} investigated the stellar mass surface density profiles of massive (M > 10$^{11.4}$ M$_\odot$) galaxies by fitting concentric ring models to HSC imaging data and performing SED fitting. Their profiles are fairly smooth compared to our own, in line with the expectations outlined in Section~\ref{fig:stacked_radial}. They find stellar mass excesses of up to 0.1 - 0.2 dex (corresponding to masses of up to 10$^{11}$ M$_\odot$ or roughly 10 percent, however in more massive galaxies than in our sample) due to low surface brightness material in the outskirts of galaxies due to low surface brightness material when comparing their profiles to cModel photometry. Their profiles for individual galaxies can extend up to 100 kpc, much further than our sample, however we stress this is probably due to our $S/N$ cut of 3 per pixel and our lower mass galaxy sample. In \citet{2018MNRAS.480..521H}, they also investigate the light profiles of these massive galaxies as a function of environment (via the halo mass), finding that similar stellar mass galaxies in more massive haloes have shallower and more extended light profiles. This can be compared to our result halo mass appears to be the main driver of the levels of merger activity we observe, whereby more massive haloes tend to display merger activity or diffuse stellar haloes.

In summary, several results we have found in our investigation here are similar to many of the results from the current literature, namely in relation to the ex-situ fraction of light or stellar mass content, as well as the radii at which this material starts to become apparent. We note as well, that all of these studies assume some sort of symmetry in their profile fitting or fitting methods, meaning, as outlined above, that the merger features are not isolated and profiles are smoothed, making a direct comparison of profiles hard to apply.

We finally postulate that although the levels of merger activity are fairly constant across our two samples, as our passive galaxy sample is older than our comparison sample, these galaxies have had more time to allow features to settle into a state of equilibrium and hence form a diffuse stellar halo, accounting for the difference in the percentage of diffuse to featureless galaxies classified in the passive galaxy and comparison samples.

\section{Conclusions}
\label{sec:conclusions}

In this work we have taken deep imaging data (down to surface brightness limits of $\mu_{\mathrm{r-band}}$ $\sim$ 28 mag arcsec$^{-2}$) of 118 low redshift, massive, central galaxies from the Subaru HSC-SSP wide survey, selected to have assembled 90 per cent of their stellar mass over 6 Gyr ago ($t_{90} > 6$ Gyr). We convolved the images with the worst FWHM of the PSFs across the five different bands, masked foreground and background objects and made cuts of $S/N > 3$ per pixel before using Voronoi binning in order to maximise the signal-to-noise in the outskirts of the galaxies. We then fitted these Voronoi bins with SEDs using \textsc{cigale} to yield stellar mass estimates. We also repeat these processes on a younger comparison sample with $t_{90} < 4$ Gyr.

Using these stellar mass estimates and the measured {\it g} and {\it r} magnitudes, we constructed radial profiles of {\it g} - {\it r} colour and stellar mass surface density. We find that the colour profiles are in good agreement with the previous work of \citet{2012MNRAS.426.2300L}. We also find expected declining stellar mass surface density profiles in the inner regions of our sample, but a flattening of the stellar mass surface density profile in the outskirts beyond $\sim 3$ R$_{\mathrm{e}}$ ($\Sigma_* \sim 10^{7.5}$ M$_\odot$ kpc$^{-2}$), driven by the low surface brightness features we observe. We find slightly bluer {\it g} - {\it r} colour profiles for the younger comparison sample (0.1 mag difference) and similar behaviour in the stellar mass density surface profiles, namely a declining profile in the central regions with excesses in the outskirts (also with $\Sigma_* \sim 10^{7.5}$ M$_\odot$ kpc$^{-2}$).

In order to compare to previous studies, \cite[cf.][]{2010ApJ...709.1018V}, we mean and median stacked and normalised all images of our galaxy sample, convolving the stacked images with a constructed PSF to investigate whether the PSF wings could make a significant difference to our profile, and thereby account for the stellar mass surface densities at larger effective radii ($\gtrsim$ 2 R$_{\mathrm{e}}$). We find the colour profiles yielded by this process in good agreement with previous studies for all of the median stacked galaxy and the PSF convolved mean and PSF convolved median stacked galaxies. We also find that the stellar mass surface density profiles are smooth and declining at all R/R$_{\mathrm{e}}$, with the PSF convolved mean stacked galaxy displaying stellar mass surface density values very similar values to previous studies (differences of 0.1 dex). We also find that accounting for the wings of the PSF by manually constructing a full PSF and convolving with the imaging data has a minimal impact on the profiles, so the PSF wings are unlikely to drive the stellar mass surface densities observed at large radii in our original profiles.

Using visual morphological classification, we also split our sample into three different categories, finding that those that display tidal features make up 42.4 per cent of our sample, similar to previous studies \citep[e.g.][]{2015MNRAS.446..120D}. Those that have no interaction signatures but display a diffuse stellar halo make up 43.2 per cent of our sample and those that are featureless make up only 14.4 per cent of our sample. When we split the profiles by these three morphological classes, we find that those classified in the features sub-sample are the drivers of the stellar mass surface densities at large effective radii ($\gtrsim$ 3 R$_{\mathrm{e}}$) with some contribution from the diffuse sub-sample. We find that our young comparison sample shows similar levels of merger activity, however many more galaxies are featureless and very few display a diffuse halo. We also find that increasing the stellar or halo mass increases the abundance of features or diffuse haloes, similar to previous studies \citep{2020MNRAS.498.2138B}.

We find that the material in these outskirts makes up a minor percentage of the total stellar mass of these systems ($\sim$ 8 per cent beyond 2 R$_{\mathrm{e}}$), corresponding to $\sim$ 10$^{10}$ M$_\odot$. This material has similar {\it g} - {\it r} colours to SDSS satellites of M$_{\mathrm{Stellar}}$ < 10$^{10.5}$ M$_\odot$, leading to a plausible scenario that this material is accreted from the surrounding satellite population.

These results show that there is an abundance of minor merger activity around central galaxies and that minor mergers can plausibly be one of the main drivers behind the size growth of massive, central galaxies.

\section*{Acknowledgements}

Thomas Jackson is a fellow of the International Max Planck Research School for Astronomy and Cosmic Physics at the University of Heidelberg (IMPRS-HD). 

The authors thank Dr. Andy Goulding for his advice and input on the use of the HSC imaging data, Dr. David Rosario for his advice on the SED fitting process and Dr. Song Huang for his advice on the known sky subtraction issues in Subaru-HSC. AP acknowledges support from the Kavli Institute for the Physics and Mathematics
of the Universe (Kavli IPMU).

The Hyper Suprime-Cam (HSC) collaboration includes the astronomical communities of Japan and Taiwan, and Princeton University. The HSC instrumentation and software were developed by the National Astronomical Observatory of Japan (NAOJ), the Kavli Institute for the Physics and Mathematics of the Universe (Kavli IPMU), the University of Tokyo, the High Energy Accelerator Research Organization (KEK), the Academia Sinica Institute for Astronomy and Astrophysics in Taiwan (ASIAA), and Princeton University. Funding was contributed by the FIRST program from Japanese Cabinet Office, the Ministry of Education, Culture, Sports, Science and Technology (MEXT), the Japan Society for the Promotion of Science (JSPS), Japan Science and Technology Agency (JST), the Toray Science Foundation, NAOJ, Kavli IPMU, KEK, ASIAA, and Princeton University. 

This paper makes use of software developed for the Large Synoptic Survey Telescope. We thank the LSST Project for making their code available as free software at  http://dm.lsst.org

This paper is based [in part] on data collected at the Subaru Telescope and retrieved from the HSC data archive system, which is operated by Subaru Telescope and Astronomy Data Center at National Astronomical Observatory of Japan. Data analysis was in part carried out with the cooperation of Center for Computational Astrophysics, National Astronomical Observatory of Japan.

\section{Data availability}

The data underlying this article are available from the public sources in the links or references given in the article (or references therein). Estimations or values from the article are available on reasonable request.

%%%%%%%%%%%%%%%%%%%%%%%%%%%%%%%%%%%%%%%%%%%%%%%%%%



\bibliographystyle{mnras}
\bibliography{lsb_paper} % if your bibtex file is called example.bib

%%%%%%%%%%%%%%%%%%%%%%%%%%%%%%%%%%%%%%%%%%%%%%%%%%

%%%%%%%%%%%%%%%%% APPENDICES %%%%%%%%%%%%%%%%%%%%%

\appendix

\section{Example Profiles}
\label{(sec:examples)}

\begin{figure*}
	\centering
	\includegraphics[width=1.8\columnwidth]{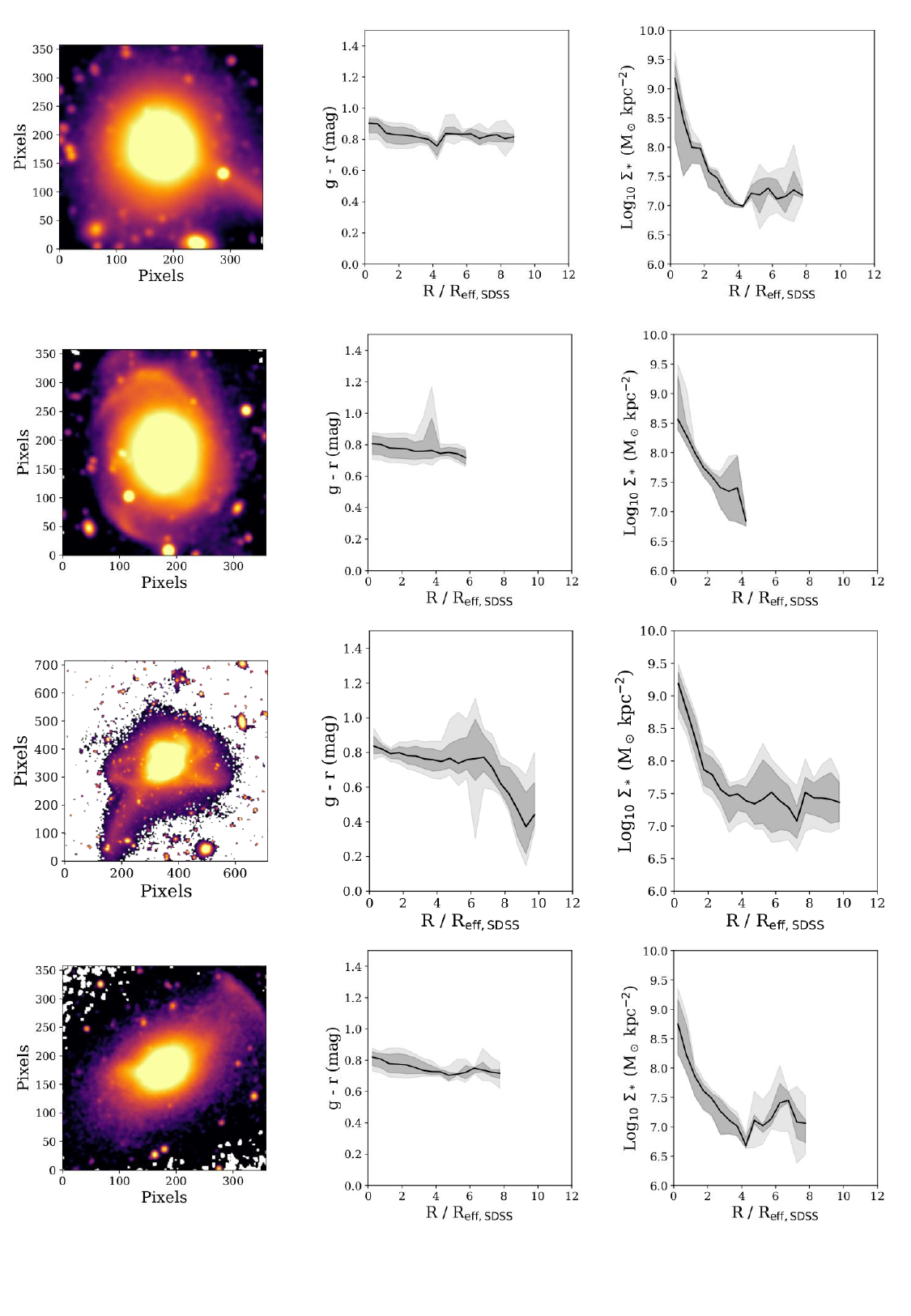}
	\caption{Example profiles of 4 HSC galaxies classified as displaying interaction features, labelled as our Features sub-sample. Each row represents one galaxy. The left-hand panel of each row displays the $i$-band image of the galaxy. The central panel represents the $g$-$r$ profile of each galaxy as a function of radius normalised to the effective radius, after detection and Voronoi binning. The right-hand shows the stellar mass surface density profiles as a function of radius normalised to the effective radius, as estimated from the SED fitting process using \textsc{cigale}.}
	\label{fig:features_profiles}
\end{figure*}

\begin{figure*}
	\centering
	\includegraphics[width=1.8\columnwidth]{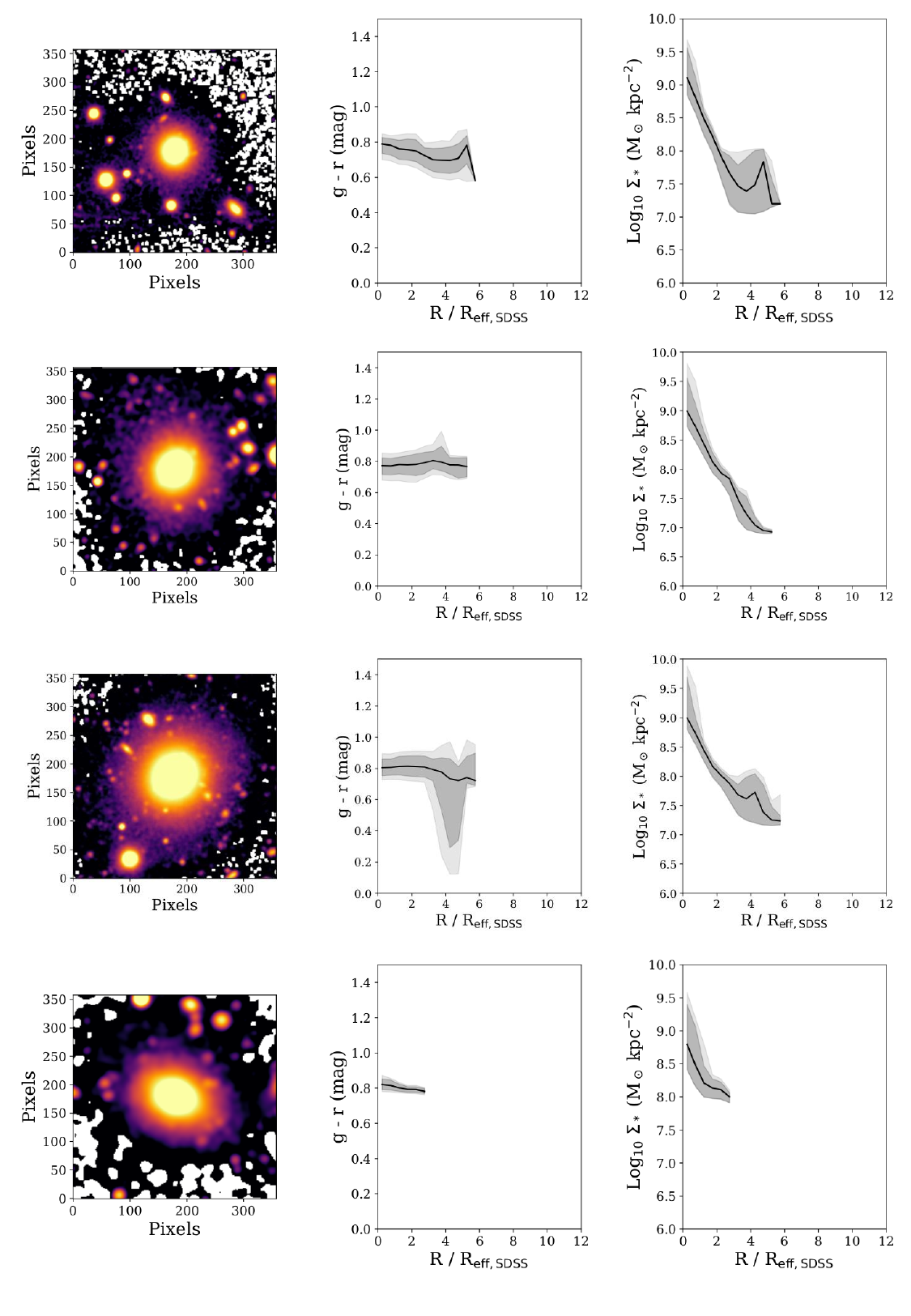}
	\caption{Example profiles of 4 HSC galaxies classified as not displaying interaction features, however displaying a diffuse stellar halo, labelled as our Diffuse sub-sample. Each row represents one galaxy. The left-hand panel of each row displays the $i$-band image of the galaxy. The central panel represents the $g$-$r$ profile of each galaxy as a function of radius normalised to the effective radius, after detection and Voronoi binning. The right-hand shows the stellar mass surface density profiles as a function of radius normalised to the effective radius, as estimated from the SED fitting process using \textsc{cigale}.}
	\label{fig:diffuse_profiles}
\end{figure*}

\begin{figure*}
	\centering
	\includegraphics[width=1.8\columnwidth]{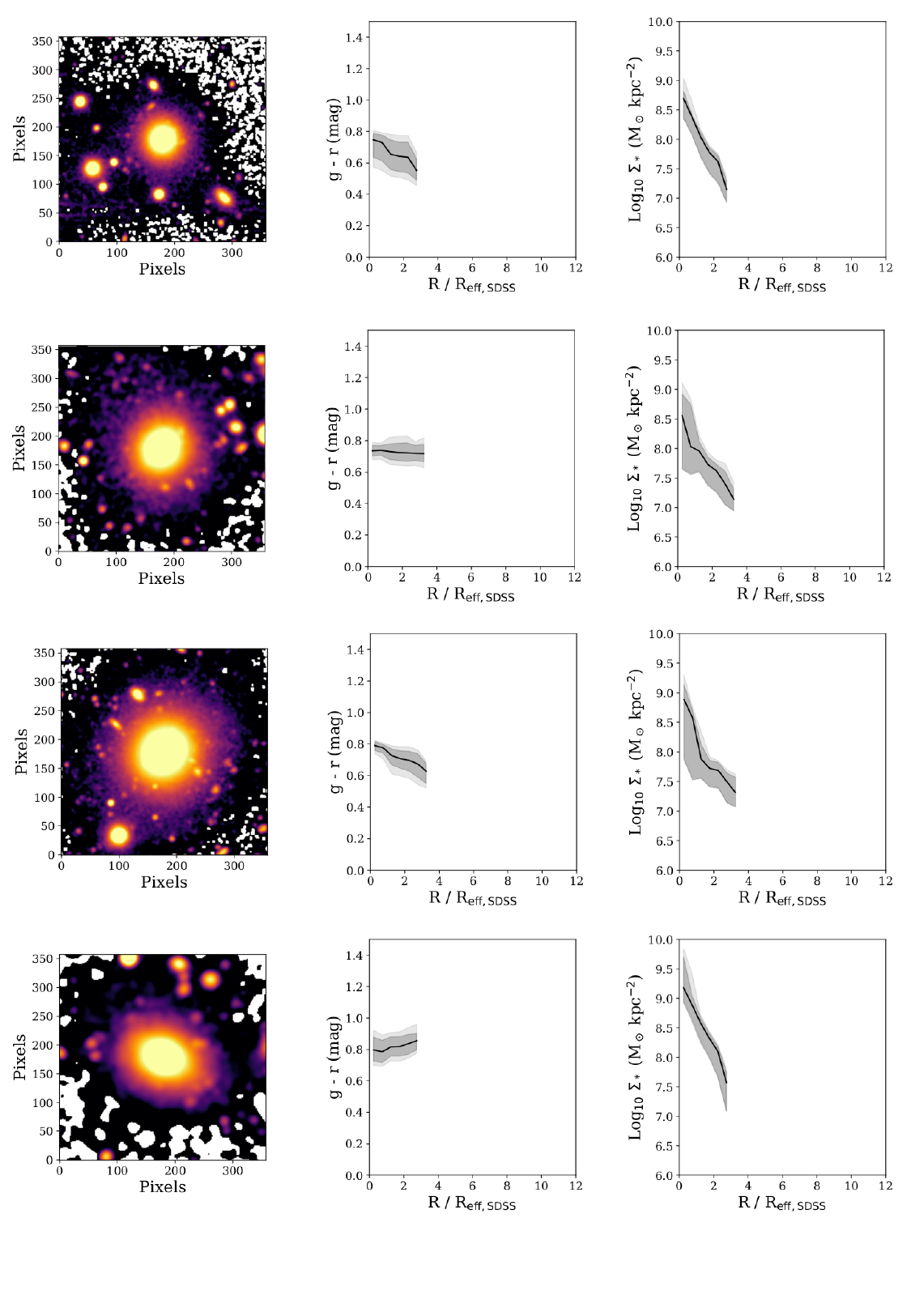}
	\caption{Example profiles of 4 HSC galaxies classified as not displaying interaction features, nor displaying a diffuse stellar halo, labelled as our Featureless sub-sample. Each row represents one galaxy. The left-hand panel of each row displays the $i$-band image of the galaxy. The central panel represents the $g$-$r$ profile of each galaxy as a function of radius normalised to the effective radius, after detection and Voronoi binning. The right-hand shows the stellar mass surface density profiles as a function of radius normalised to the effective radius, as estimated from the SED fitting process using \textsc{cigale}.}
	\label{fig:featureless_profiles}
\end{figure*}

%%%%%%%%%%%%%%%%%%%%%%%%%%%%%%%%%%%%%%%%%%%%%%%%%%

% Don't change these lines
\bsp	% typesetting comment
\label{lastpage}
\end{document}